%% file: main.tex
\documentclass[usenames,dvipsnames]{aa_edited}

\usepackage{graphicx,txfonts,booktabs,natbib,mathtools,microtype,amssymb,amsmath,xspace,siunitx,physics,nicefrac,comment,nicematrix,multirow}
\usepackage{orcidlink}
\definecolor{linkcolor}{rgb}{0.0,0.3,0.5}
 \hypersetup{
     colorlinks=true,
     linkcolor=linkcolor,
     filecolor=linkcolor,
     citecolor = linkcolor,      
     urlcolor=linkcolor,
     }

\newcommand{\mrs}{Aix-Marseille Universit\'e, Universit\'e de Toulon, CNRS, CPT, Marseille, France}
\newcommand{\milan}{Dipartimento di Fisica ``G. Occhialini'', Universit\'a degli Studi di Milano-Bicocca, Piazza della Scienza 3, 20126 Milano, Italy}
\newcommand{\infn}{INFN, Sezione di Milano-Bicocca, Piazza della Scienza 3, 20126 Milano, Italy}

\graphicspath{figures/} 

\usepackage{txfonts}
\begin{document} 
\title{Cosmology with the angular cross-correlation of gravitational-wave and galaxy catalogs: Forecasts for next-generation interferometers and the Euclid survey}
   \titlerunning{A\&A, 712, A37 (2026)}
   \author{Alessandro Pedrotti\inst{1}\fnmsep\thanks{Corresponding author: alessandro.PEDROTTI@univ-amu.fr}$\,$\orcidlink{0009-0005-8389-6678}
          \and
          Michele Mancarella\inst{1}$\,$\orcidlink{0000-0002-0675-508X}
          \and Julien Bel\inst{1}$\,$\orcidlink{0009-0006-7837-1866}
          \and Michele Santoni\inst{1}$\,$\orcidlink{0009-0009-3992-4749}
           \and Davide Gerosa\inst{2,3}$\,$\orcidlink{0000-0002-0933-3579}
          }
    \authorrunning{Pedrotti, A., et al.}
    \institute{\mrs
         \and
             \milan
             \and
             \infn
             }
   \date{}

 
  \abstract
{The spatial clustering of galaxies has long been a key probe of cosmology. Gravitational–wave (GW) sources, which provide direct luminosity–distance measurements, have recently emerged as a complementary tracer of large–scale structures. The cross–correlation of GW and galaxy catalogs offers a novel way to test cosmic expansion.}
{We investigate the potential of tomographic GW–galaxy angular power spectra to constrain cosmological parameters, focusing on the Hubble constant and matter density, in the context of third–generation (3G) GW detectors combined with the Euclid survey.}
{We constructed our forecasts using realistic GW source populations and error models calibrated on recent detector designs. We adopted a Fisher–matrix approach, marginalized over nuisance parameters including tracer biases, primordial spectrum parameters, and baryon density, and compared different survey configurations, binning schemes, and GW detector networks.}
{We find that tomographic cross–correlation can constrain $H_0$ at a percent or sub–percent precision, depending on the binning strategy, network configuration, and observing time. Combining galaxy autocorrelations with GW–galaxy cross–correlations improves constraints by up to a factor of ${\sim}10$ relative to either probe alone. We further show that this performance requires multiple interferometers with accurate sky localization, and we discuss the added value of spectroscopic surveys and the detectability of GW clustering bias.}
{Our results demonstrate that this technique, applied to 3G GW detectors in synergy with large galaxy surveys, can deliver competitive measurements of cosmic expansion, even when marginalizing over a wide range of astrophysical and cosmological nuisance parameters.}


   \maketitle
\makeatletter
\fancyhead[CE]{\aa@headfont Pedrotti, A., et al.: A\&A, 712, A37 (2026)}
\makeatother
%

\section{Introduction}

The study of the three–dimensional large–scale distribution of galaxies is a cornerstone of modern cosmology. Its statistical properties probe structure formation, cosmology, and gravity~\citep{Kaiser:1987qv,Szalay:1997cc,Yoo:2009au,Bonvin:2011bg,Challinor:2011bk}, and are mapped by redshift surveys such as the Dark Energy Spectroscopic Instrument (DESI)~\citep{DESI:2016fyo}, Euclid~\citep{EUCLID:2011zbd,Euclid:2024yrr}, the Legacy Survey of Space and Time (LSST)~\citep{LSSTScience:2009jmu}, and the Square Kilometre Array (SKA)~\citep{Braun:2019gdo,Bull:2015lja}. The DESI survey recently released the latest clustering results~\citep{DESI:2024hhd}, while Euclid will soon provide analyses of two–point correlations~\citep{Euclid:2025gde}, the power spectrum~\citep{Euclid:2023tog,Euclid:2024ufa}, and lensing–clustering combinations~\citep{Euclid:2024fsk,Euclid:2023qyw,Euclid:2023pyq,Euclid:2024xqh}. Specifically, Euclid will deliver both a spectroscopic sample of $\sim 10^7$ galaxies at $0.9<z<1.8$ and a photometric sample of $\sim 10^9$ galaxies up to $z\sim2$~\citep{Euclid:2024yrr}.

While galaxies are an established observational target for mapping the Universe, gravitational waves (GWs) have emerged in recent years as a new, complementary tracer.
Since their first direct detection in 2015~\citep{LIGOScientific:2016aoc}, the LIGO-Virgo-Kagra (LVK) collaboration has reported 218 candidate detections from compact binary mergers~\citep{LIGOScientific:2018mvr,LIGOScientific:2020ibl,KAGRA:2021vkt,LIGOScientific:2025slb}.
These sources are complementary to galaxies because they provide a direct measurement of the luminosity distance to the source~\citep{Schutz:1986gp}, but no direct redshift measurement~\citep{Schutz:1986gp,Finn:1992xs}.
This naturally motivates the study of GW clustering in distance space~\citep{Namikawa:2015prh,Namikawa:2016edr,Zhang:2018nea,Namikawa:2020twf,Libanore:2020fim,Garoffolo:2020vtd,Kumar:2022wvh,Zheng:2023ezi,Vijaykumar:2020pzn,Zazzera:2023kjg,Begnoni:2024tvj,Dehghani:2024wsh}, recently formulated in full generality by~\cite{Fonseca:2023uay} and shown to differ from its analog in redshift space~\citep{Bonvin:2011bg}.
Because compact binaries form inside galaxies, both populations trace the \emph{same} underlying dark matter field (with possible exceptions for primordial black holes\footnote{Even in this case, they remain correlated with the dark matter distribution, though with a different amplitude~\citep{Raccanelli:2016cud,Scelfo:2018sny}.}). This enables the use of two–point GW–galaxy cross–correlations for both astrophysics~\citep{Raccanelli:2016cud,Scelfo:2018sny,Scelfo:2020jyw,Calore:2020bpd,Scelfo:2021fqe,Libanore:2021jqv,Balaudo:2023klo,Gagnon:2023mnd,Zazzera:2024agl} and cosmology.

The key cosmological feature is that the two tracers are observed in different spaces: galaxies in redshift space and GWs in distance space. Their cross–correlation in tomographic redshift and distance bins is therefore maximized only along the correct distance–redshift relation, allowing constraints on its parameters and, in particular, on the Hubble constant, $H_0$. First suggested by~\cite{Oguri:2016dgk},~\cite{Ferri:2024amc} recently studied this in more detail. Other approaches based on two-point information in cosmology have been proposed~\citep{Camera:2013xfa,Nair:2018ign,Mukherjee:2019wcg,Bera:2020jhx,Mukherjee:2020hyn,Mukherjee:2020mha,Diaz:2021pem,Mukherjee:2022afz,Ghosh:2023ksl,Beltrame:2024cve}. These approaches are complementary to dark siren methods, which rely on prior information on the GW source population (see~\citealp{Pierra:2025fgr}, and references therein for a recent review). Importantly, cross-correlation techniques involve distinct systematics that remain to be fully explored, which makes them a complementary probe that can be used alongside other approaches.

Crucially, this program will reach its full statistical power with third–generation (3G) GW observatories: the Einstein Telescope (ET) in Europe~\citep{ET:2025xjr} and the Cosmic Explorer (CE) in the US~\citep{Evans:2023euw}. These facilities are expected to detect millions of sources over ${\sim}10$ years~\citep{Iacovelli:2022bbs,Branchesi:2023mws}, comparable to galaxy survey catalogs, with subdegree angular resolution~\citep{Borhanian:2022czq,Iacovelli:2022bbs,Branchesi:2023mws,Gupta:2023lga}. Third-generation (3G) detectors aim to achieve percent–level cosmological measurements of $H_0$, which is critical to address the current tension between supernovae (SNe) and cosmic microwave background (CMB)~\citep{Riess:2021jrx,Planck:2018vyg}, and to probe possible deviations from the standard cosmological model.
These facilities are currently in a critical stage of planning. In particular, the ET collaboration is currently evaluating designs and sites~\citep{Branchesi:2023mws,ET:2025xjr}. Synergies with CE are expected to be particularly powerful for cosmology and multimessenger science.

In this work, we assess the potential of 3G GW detectors to constrain the expansion history of the Universe through the GW–galaxy two–point correlation in tomographic redshift bins. We place particular emphasis on the Euclid survey. Cosmological studies based on cross-correlation techniques have been developed for LVK detectors, typically targeting $H_0$~\citep{Mukherjee:2019wcg,Bera:2020jhx,Mukherjee:2020hyn,Diaz:2021pem,Mukherjee:2022afz,Ghosh:2023ksl}. In contrast, the 3G literature so far has focused mainly on astrophysical applications such as clustering bias measurements~\citep{Raccanelli:2016cud,Scelfo:2018sny,Scelfo:2020jyw,Calore:2020bpd,Scelfo:2021fqe,Libanore:2021jqv,Zazzera:2024agl}.~\cite{Ferri:2024amc} recently presented a dedicated investigation of constraints on cosmic expansion using the cross–correlation technique of~\cite{Oguri:2016dgk}. Their analysis is based on simulations with fixed tracer bias and cosmological parameters, except for those governing the distance-redshift relation.
In this work we take a complementary approach. Using a Fisher matrix formalism, we forecast the statistical power of 3G data, including nuisance parameters. Specifically, we adopt a fully agnostic treatment of tracer bias and marginalize over additional cosmological parameters. This allows us to explicitly quantify the information gain from combining galaxy autocorrelations with GW–galaxy cross–correlations under realistic assumptions.
Furthermore, we employ survey specifications and binning schemes calibrated to official Euclid requirements, and compare the constraining power of the photometric and spectroscopic samples. In the latter case, we also compare to a more optimistic design, namely the SKA Phase II spectroscopic survey~\citep{Bull:2015lja}.
We systematically investigate different binning strategies and quantify their impact on parameter inference. On the GW side, we adopt updated sensitivity curves for ET and CE and consider multiple 3G network configurations. We explicitly compare alternative ET geometries and their synergies with CE, with realistic observational errors on distance and sky localization. We also use updated compact–binary population models and revisit the detectability of the GW clustering bias under model-independent assumptions.
Altogether, our forecasts provide a comprehensive assessment of how 3G detectors, in synergy with Euclid, can deliver competitive and robust constraints on the expansion history while accounting for astrophysical and cosmological uncertainties.

\section{Methodology}

\subsection{Observable and cross-correlation signals}
\begin{figure*}[t]
  \centering
    \includegraphics[width=0.9\textwidth]{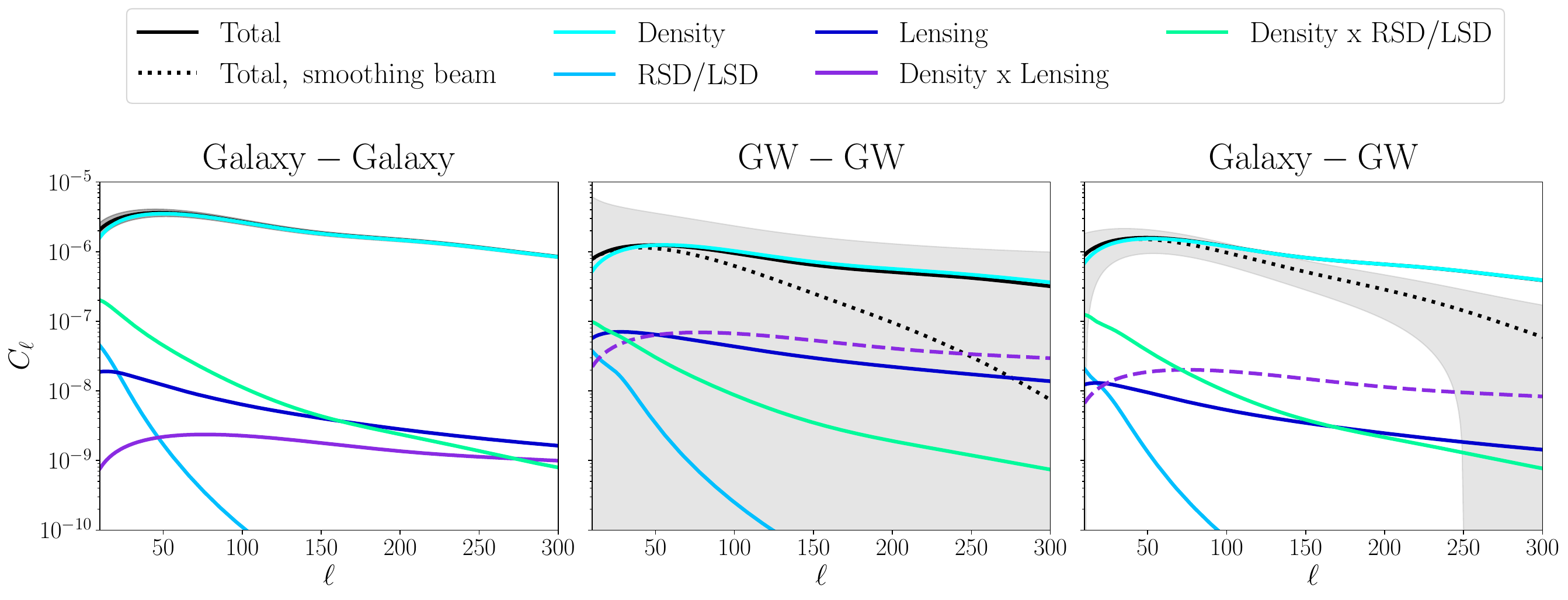}
  \caption{Angular power spectra for the auto- and cross-correlation of the Euclid photometric sample with a GW catalog from five years of observations with two ET (L-shaped) and two CE detectors, shown for the redshift bin $z\in[1,1.1]$. Colors indicate the different contributions to number–count fluctuations (see legend). Solid lines assume perfectly localized GW sources, while dotted black lines in the central and right panels include realistic smoothing due to their imperfect localization. Shaded areas show $1\sigma$ errors. Dashed lines indicate negative terms, which are plotted on the positive axis for comparison.}
  \label{fig:signal}
\end{figure*}
We considered the number counts of tracers $X,Y$ of the underlying dark matter field (here, galaxies and GWs) as a function of sky position $\mathbf{n}$ and a tracer–dependent radial coordinate $x$ (redshift $z$ for galaxies, luminosity distance $d_L$ for GWs). The formalism for cross–correlations is given in~\cite{Scelfo:2018sny} (see also~\citealp{Raccanelli:2016cud,Scelfo:2020jyw,Scelfo:2021fqe,Scelfo:2022lsx,Bosi:2023amu,Namikawa:2020twf,Fonseca:2023uay,Libanore:2020fim,Libanore:2021jqv}).

The fractional fluctuation of the tracer $X$ at $(\mathbf{n},x)$ is
\begin{equation}\label{eq:Delta_def}
\Delta^X(\mathbf{n},x) = \frac{N^X(\mathbf{n},x)-\langle N^X\rangle(x)}{\langle N^X\rangle(x)} \, .
\end{equation}
This is a stochastic field with zero mean, and can be expanded in spherical harmonics, i.e.,
\begin{equation}\label{eq:slm_def}
\Delta^X(\mathbf{n},x)=\sum_{\ell=0}^{\infty}\sum_{m=-\ell}^{\ell}s^X_{\ell m}(x)Y_{\ell m}(\mathbf{n}) \, . 
\end{equation}
The statistical properties of the field are encoded in the angular power spectrum in tomographic bins $C_{\ell}^{XY}(x_i, x_j)$, obtained by cross--correlating the harmonic coefficients of tracer $X$ in bin $i$ (centered at the radial coordinate $x_i$) with those of tracer $Y$ in bin $j$ (centered at the radial coordinate $x_j$):
\begin{equation}
\langle s^{X}_{\ell m}(x_i) s^{Y^*}_{\ell' m'}(x_j) \rangle = \delta_{\ell \ell'} \delta_{m m'} C_{\ell}^{XY}(x_i, x_j) \, .
\end{equation}
The explicit expression of the power spectrum follows from the galaxy number density fluctuation in perturbation theory, binned into weighted redshift and distance intervals. This can be written as\footnote{We adopted the \texttt{class} convention for the normalization of Fourier transforms, see e.g., Ref.~\cite{DiDio:2013bqa}.}
\begin{equation} \label{angular_ps}
C_\ell^{XY}(x_i,x_j)=\frac{2}{\pi}\int \dd k \, k^2 \, P(k) \,\Delta_\ell^{X,x_i}(k) \,\Delta_\ell^{Y,x_j}(k) \, ,
\end{equation}
where $P(k)$ is the primordial spectrum, and
\begin{equation}\label{eq:source_num_den}
\Delta_\ell^{X,x_i}(k) = \int_{0}^{\infty} \dd x \, w^X(x,x_i) \, \Delta_\ell^X(k,x) 
\end{equation}
is the effective transfer function of the tracer $X$. The weight is a normalized window function centered at coordinate $x_i$. This can be expressed in terms of an unweighted window function $W^X(x,x_i)$ (encoding how the observed events are binned in either redshift or distance; more details are provided in Sect.~\ref{sec:noise_and_err}) and the observed distribution of tracer X, $ \dd N^X_{\rm obs}/\dd x$, as follows:
\begin{equation}\label{window}
w^X(x,x_i) = \frac{W^X(x,x_i) \,\dd N^{X}_{\rm obs}/\dd x}{\int \dd x'\, W^X(x',x_i) \,\dd N^{X}_{\rm obs}/\dd x'} \, ,
\end{equation}
The kernel $\Delta_\ell^X(k,x)$ encodes relativistic number–count contributions from perturbation theory.~\cite{Bonvin:2011bg} and~\cite{Challinor:2011bk} derived this expression in redshift space ($x\equiv z$);~\cite{Zhang:2018nea} and~\cite{Libanore:2020fim} first studied it in the luminosity–distance space for GWs and SNe ($x\equiv d_L$); while~\cite{Namikawa:2020twf} and~\cite{Fonseca:2023uay} provided a general derivation. The results differ from their counterparts in redshift space.
Contributions are grouped into density, velocity (redshift-space distortion (RSD) + Doppler), lensing, and gravity terms. In luminosity–distance space, the RSD contribution becomes a ``luminosity distance space distortion'' (LSD) term~\citep{Zhang:2018nea,Libanore:2020fim,Fonseca:2023uay}.
This can be written schematically as
\begin{equation}\label{relC}
\begin{split}
       \Delta_\ell^X(k,x) = &\Delta_\ell^{X, {\rm den}}(k,x)+\Delta_\ell^{X, {\rm len}}(k,x)+\Delta_\ell^{X, {\rm RSD/LSD}}(k,x)+\\
       &\Delta_\ell^{X, {\rm dop}}(k,x)+\Delta_\ell^{X, {\rm gr}}(k,x) \, .
\end{split}
\end{equation}
Here, we retain the dominant terms, namely density, lensing, and RSD/LSD (see~\cite{Fonseca:2023uay} for a full comparison)\footnote{%
Although the signal is dominated by the density term, other contributions should be modeled since they affect inter-bin covariances and thus the forecasted errors, particularly the lensing term~\citep{Scelfo:2018sny}.}.
We computed Eq.~\ref{angular_ps} using the Limber approximation~\citep{Limber:1954zz,LoVerde:2008re}, restricting to scales where it is valid (Sect.~\ref{sec:summary}). Appendix~\ref{app:rel_numb_count_expr} details the explicit formulae.

With this formalism, we constructed the auto- and cross-correlation spectra of two given tracers. Specifically, we studied the galaxy-galaxy and GW-GW autocorrelation terms, $C_\ell^{g-g}$ and $C_\ell^{\rm GW-GW}$, respectively, and their cross--correlation $C_\ell^{\rm GW-g}$.
Figure~\ref{fig:signal} shows these spectra for the Euclid photometric survey and a GW detector network with two ET and two CE detectors in a redshift bin $z\in[1,1.1]$ (survey details in Sect.~\ref{sec:simulations}). Colored lines indicate density, RSD, LSD, and lensing contributions, as well as their cross–terms\footnote{RSD/LSD–lensing cross–spectra are subdominant and omitted from the figure, but included in the analysis.}. Solid lines assume perfect GW localization, while dotted lines include smoothing due to limited sky localization (Sect.~\ref{sec:noise_and_err}). The shaded bands give the expected $1\sigma$ errors.

\subsection{Cosmological information and dependence on population properties and selection effects}\label{sec:pop_pars}
\begin{figure}[t]
  \centering    \includegraphics[width=0.9\columnwidth]{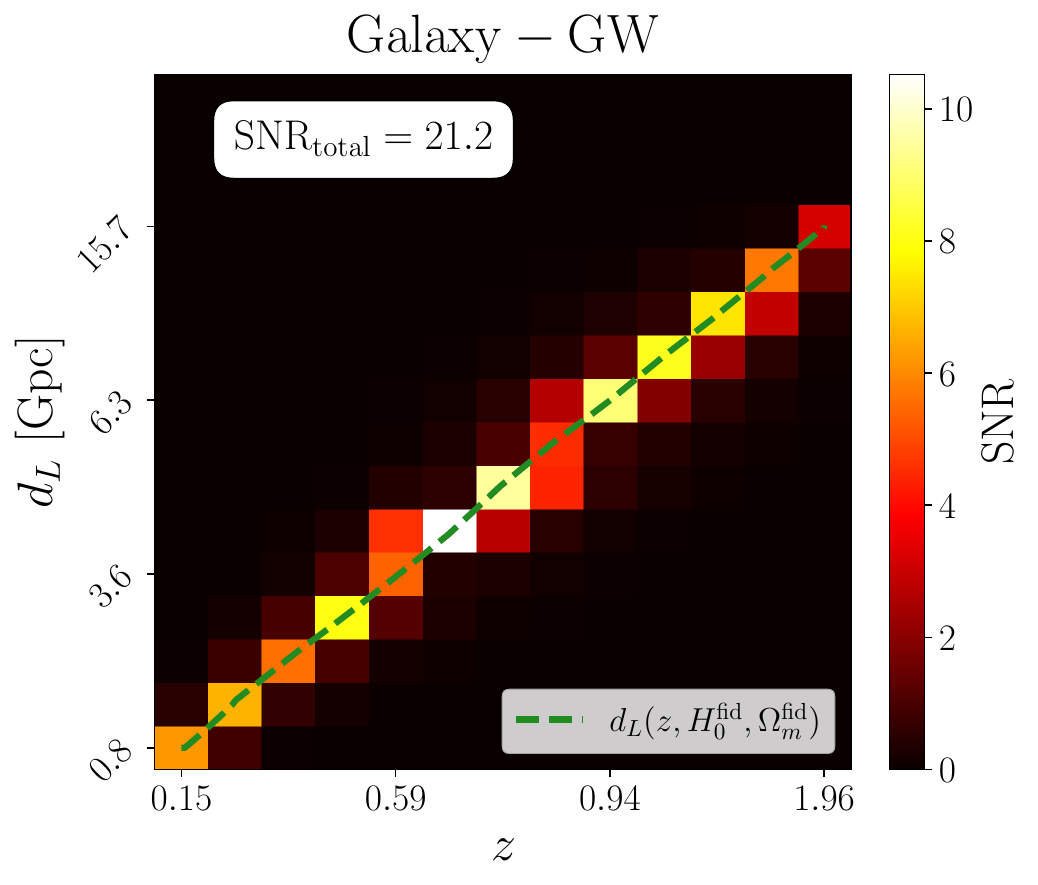}
  \caption{Signal-to-noise ratio (S/N) of the cross–correlation between the Euclid photometric sample and a GW catalog from a network of two L-shaped ET detectors and two CE detectors. The S/N peaks along the fiducial distance–redshift relation (dashed green).}
  \label{fig:SNR}
\end{figure}

Examination of Eq.~\ref{angular_ps} clarifies the source of cosmological information contained in the tomographic cross–spectrum $C_\ell^{\rm GW-g}$. The dominant contribution is the density term, which relates the number-count fluctuations to the matter density through the tracer bias. In the Limber approximation, the expression is (see Appendix~\ref{app:rel_numb_count_expr}):
\begin{equation} \label{C_l limber}
\begin{split}
C_\ell^{\rm GW-g}(z_i, d_{L, j}) = & \int \dd z \, w^{\rm g}(z , z_{i}) \, w^{\rm GW}[ d_L(z,\lambda), d_{L, j}] \,\frac{\dd d_L}{\dd z}(z, \lambda) \\
& \times \frac{H(z)}{c \, r(z)^2} \, b_{\rm GW}(z) \, b_{\rm g}(z) \, P\left[\tfrac{\ell+1/2}{r(z, \lambda)},z\right] \, ,
\end{split}
\end{equation}
where the tracer biases are $b_{\rm g}$ and $b_{\rm GW}$, the comoving distance is $r(z,\lambda)=d_L(z,\lambda)/(1+z)$, and $\lambda$ denotes the parameters of the distance–redshift relation. Eq.~\ref{C_l limber} shows that if the galaxy window function $w^{\rm g}$ has non-zero support in a bin around redshift $z_i$, the GW window function $w^{\rm GW}$ will be non-vanishing in the (fixed) bin around $d_{L, j}$ only if $d_L(z,\lambda)$, for redshifts around $z_i$, falls inside the $d_{L, j}$ bin. That is, the cross–term is maximized when the GW and galaxy bins align along the correct distance–redshift relation~\citep{Oguri:2016dgk}. Figure~\ref{fig:SNR} illustrates this using the signal-to-noise-ratio (S/N) of the cross-correlation between the Euclid photometric sample and a GW detector network composed of two ET and two CE detectors. We computed the S/N using Eqs.~\ref{snr_ell}--\ref{snr_cumulative} below, following~\cite{Scelfo:2020jyw}. We divided the Euclid sample into 13 equally-populated redshift bins. We used a Planck 2018 cosmology to convert redshift bins to distance, which defined the binning of the GW survey\footnote{We included three extra bins at the high-distance edge to fully cover the observed sample.}.
The S/N peaks along the true distance–redshift relation (dashed green line). This property allows to constrain its parameters~\citep{Oguri:2016dgk}.

Besides $\lambda$, the signal depends on cosmological parameters $({A_s,n_s,\Omega_{\rm b}h^2})$ that enter the power spectrum and on tracer properties. These must be marginalized over to avoid underestimating the uncertainty and bias.
Four key terms encode the dependence on population properties~\citep{Scelfo:2020jyw}:
\begin{itemize}
\item \emph{The observed redshift and distance distributions,} $\dd N^X_{\rm obs}/\dd x$. This sets the weighting factor in the window function (Eq.~\ref{window}). This can be determined directly from the observed sample. The cross–correlation then compares \emph{fluctuations} in GW and galaxy counts, whose statistics are insensitive to the overall distribution or selection function. Different selection cuts or variations in the background source distribution alter the clustering of the remaining sources, i.e., their bias (next point). Hence, the assumed form of the binary black hole (BBH) merger distribution is likely not the dominant source of systematic uncertainty for this approach, although its impact has not yet been studied in detail and should be explicitly assessed in future work. By contrast, the clustering bias is certainly a possible source of systematic error and must therefore be modeled robustly, either by comprehensive forward modeling or by adopting a non-parametric approach, as we do in this work.

\item \emph{Clustering bias,} $b_{X}$. Source counts trace the dark matter density through a bias term (Appendix~\ref{app:rel_numb_count_expr}). On the large scales considered here, the bias depends only on the redshift~\citep{Fosalba:2013mra}. Varying the bias linearly changes the signal amplitude. If the bias evolves with redshift, it can introduce correlations with the cosmological parameters. Here, we include a free bias parameter for each redshift bin and tracer. This enlarges the parameter space, but avoids strong assumptions on the functional form.

\item \emph{Magnification bias and local count slope,} $s_{X}$. This encodes lensing–induced changes in the number counts near the detection threshold and is defined as the derivative of the comoving density with respect to the threshold. It affects lensing (Appendices~\ref{app:rel_numb_count_expr}, \ref{app:popdetails}) and, more weakly, gravity and Doppler terms~\citep{Alonso:2015uua,Maartens:2021dqy,Fonseca:2023uay,Zazzera:2023kjg}.

\item \emph{Evolution bias, $b_{{\rm e}, X}$.} This encodes variations in source counts from the formation of new objects at a fixed detection threshold. Similarly to magnification bias, it depends on both populations and survey specifications. In redshift space it enters the gravity term and, weakly, the velocity term, so we neglect it here. In distance space, it also affects lensing~\citep{Fonseca:2023uay} (Appendix~\ref{app:rel_numb_count_expr}).
\end{itemize}
Since the evolution and magnification bias enter only in subdominant terms, we fix them in this work.

\subsection{Noise and measurement uncertainties}\label{sec:noise_and_err}

In addition to the angular power spectrum, we considered the effect of shot noise, which we assumed to be uncorrelated in different experiments and redshift and distance bins.
This can also be decomposed into harmonic coefficients $n_{\ell m}(x_i)$. We obtain the harmonic coefficients of the observed signal by summing the signal and noise,
\begin{equation}
    a^X_{\ell m}(x_i) = s_{\ell m}(x_i) + n_{\ell m}(x_i) \, .
\end{equation}
The expectation value of the noise is given by
\begin{equation}
 \langle n^{X}_{\ell m}(x_i) n^{Y^*}_{\ell' m'}(x_j) \rangle = \delta_{\ell \ell'} \delta_{m m'} \delta_{X Y} \delta_{ij} 
 \mathcal{N}^X(x_i) \, ,
\end{equation}
where  
\begin{equation}\label{shot}
 \mathcal{N}^X(x_i)  = {\left[ \int_0^{\infty} \! \dd x \, W^X(x,x_i)\,\frac{\dd^2 N^{X}_{\rm obs}}{\dd x\, \dd \Omega} \right]}^{-1}
\end{equation}
is the inverse total number of observed sources per steradian in the $i$-th bin\footnote{We note that $\mathcal{N}^X(x_i)$ depends on the overall source merger rate rather than just the shape of the distribution, as is the case for the signal $C_\ell$.}.
We also assumed that signal and noise are uncorrelated, i.e.,
\begin{equation}
 \langle n^{X}_{\ell m}(z_i) s^{Y^*}_{\ell' m'}(z_j) \rangle = 0 \, .
\end{equation}
Using these properties, the observed angular power spectrum $\tilde{C}_\ell^{XY}$ can be written as 
\begin{equation} \label{Cl_tilde}
\begin{split}
    & \langle a_{\ell m}^X(x_i), a^{Y^*}_{\ell' m'}(x_j) \rangle = \delta_{\ell \ell'}\delta_{m m'}\tilde{C}_\ell^{XY}(x_i,x_j) \, , \\
    & \tilde{C}_\ell^{XY}(x_i,x_j) \equiv C_\ell^{XY}(x_i,x_j)+\delta_{XY}\delta_{ij}\mathcal N^X(x_i) \, .
    \end{split}
\end{equation}

For GW sources, we accounted for the limited angular resolution of the detectors. The best achievable resolution sets a limiting scale in the angular power spectrum~\citep{Oguri:2016dgk,Libanore:2020fim,Scelfo:2022lsx,Bosi:2023amu}, given by $\ell_{\rm max} = \pi/\Delta \Omega_{\rm min}^{1/2}$, where $\Delta\Omega_{\rm min}$ denotes the best $1\sigma$ sky–localization of the GW catalog. We imposed a hard cut on the signal at the limiting scale $\ell_{\rm max}$, and additionally modeled the suppression due to imperfect localization as an exponential damping. This follows from treating the localization error as a Gaussian beam and convolving it with the GW density contrast~\citep{Namikawa:2016edr,Calore:2020bpd,Libanore:2020fim,Scelfo:2022lsx,Bosi:2023amu}. Thus, whenever one of the two tracers is a GW source, one has
\begin{align}
& C_\ell^{\rm GW-gal}(d_{L,i},z_{j})  \mapsto C_\ell^{\rm GW-gal}(d_{L,i},z_{j}) \times e^{ \nicefrac{-\ell(\ell+1)}{\ell^2_{\rm damp}} } \, , \label{shotGW} \\
& C_\ell^{\rm GW-GW}(d_{L,i},d_{L,j})  \mapsto C_\ell^{\rm GW-GW}(d_{L,i},d_{L,j}) \times e^{ \nicefrac{-2 \ell(\ell+1)}{\ell^2_{\rm damp}} } \, , \\
& \ell^2_{\rm damp}  = \frac{(2\pi)^{3/2}}{\Delta\Omega_{1\sigma}} \, ,
\end{align}
where $\Delta\Omega_{1\sigma}$ is the solid angle corresponding to the $68 \%$ contour level of the angular error\footnote{Slightly different definitions of the damping factors appear in the literature, yielding different values of the numerical coefficient in the definition of $\ell^2_{\rm damp}$. We are not aware of any established derivation of this numerical factor, so we provide one in Appendix~\ref{app:multipoles}.}.

Finally, the error on distance or redshift can be included in the unweighted window function by modeling it as the convolution between a ``bare'' window function and the likelihood of the distance or redshift profile~\citep{Oguri:2016dgk,Calore:2020bpd}.
Therefore, for both GWs and galaxies, we modeled the unweighted window function $W^X(x,x_i)$ appearing in Eq.~\ref{window} as a step function $S^X(x,x_i)$ convoluted with a log-normal likelihood encoding the measurement uncertainty, denoted by $\mathcal{L}(x^{\rm obs}|x)$.
This yields
\begin{equation}\label{window_with_err}
\begin{split}
    W^X(x,x_i) = & \int_0^\infty \dd y \, S^X(y, x_i)\,\mathcal{L}(y|x)=\frac{1}{2}\Big\{\mathrm{erf}\big[u(x^{i+1}, x)\big]- \\
    &\mathrm{erf}\big[u(x^i, x)\big]\Big\}\, , \\
    u(y, x)= & \frac{\ln y -\ln x }{\sqrt{2}\sigma_{\ln x}} \, ,
    \end{split}
\end{equation}
where $\sigma_{\ln x}$ is the standard deviation of the log-normal, that is, the relative error on redshift or distance, and $x^i$, $x^{i+1}$ denote the lower and upper edges of the $i$th bin in distance or redshift.

\subsection{Fisher matrix and S/N}\label{sec:fisher}

We estimated measurement errors using the Fisher matrix formalism.
For a set of parameters $\{\theta_\alpha\}$, the Fisher matrix for the angular power spectrum assuming a Gaussian likelihood for the harmonic coefficients $a_{\ell m}$ is~\citep{Bellomo:2020pnw,Abramo:2022qir}
\begin{equation} \label{fisher_mat}
F_{\alpha \beta} = -\left\langle \frac{\partial^2 \ln \mathcal{L}}{\partial \theta_\alpha \partial \theta_\beta} \right\rangle= f_{\rm sky}\sum_\ell\frac{2\ell+1}{2}\mathrm{Tr}[\mathcal{C}_\ell^{-1}(\partial_\alpha\mathcal{C}_\ell)\mathcal{C}_\ell^{-1}(\partial_\beta\mathcal{C}_\ell)]
\end{equation}
where $f_{\rm sky}$ is the fraction of the sky observed by the experiment and $\mathcal{C}_\ell$ is the covariance matrix. 
Specifically, we considered a galaxy survey binned into $N$ redshift bins $z_1 , ..., z_N$, and a GW catalog binned into $M$ luminosity distance bins $d_{L,1} , ..., d_{L,M}$. The (symmetric) covariance matrix $\mathcal{C}_\ell$ is expressed as~\citep{Scelfo:2018sny}
\begin{equation}
\resizebox{0.95\columnwidth}{!}{$
\mathcal{C}_\ell =
\begin{pmatrix}
\tilde{C}_{\ell}^{gg}(z_1, z_1) & \cdots & \tilde{C}_{\ell}^{gg}(z_1, z_N) & \tilde{C}_{\ell}^{gGW}(z_1, d_{L,1}) & \cdots & \tilde{C}_{\ell}^{gGW}(z_1, d_{L,M}) \\
 & \ddots & \vdots & \vdots & & \vdots\\
 & & \tilde{C}_{\ell}^{gg}(z_N, z_N) & \tilde{C}_{\ell}^{gGW}(z_N, d_{L,1}) & \cdots & \tilde{C}_{\ell}^{gGW}(z_N, d_{L,M}) \\
 & & & \tilde{C}_{\ell}^{GWGW}(d_{L,1}, d_{L,1}) & \cdots &  \tilde{C}_{\ell}^{GWGW}(d_{L,1}, d_{L,M})\\
 & & & & \ddots & \vdots\\
 & & & & & \tilde{C}_{\ell}^{GWGW}(d_{L,M}, d_{L,M})\\
\end{pmatrix}
$}
\end{equation}
where $\tilde{C}_{\ell}$ is the observed angular power spectrum as in Eq.~\ref{Cl_tilde}. 
The matrix $\partial_\alpha \mathcal{C}_\ell = \partial\mathcal{C}_\ell/\partial\theta_\alpha$ contains the derivatives of the covariance-matrix elements with respect to a given cosmological parameter $\theta_\alpha$, while keeping all other parameters fixed.
The forecasted marginal error on the parameter $\theta_\alpha$ is given by the square root of the diagonal element of the inverse Fisher matrix, $\sigma_{\theta_\alpha}=\sqrt{F^{-1}_{\alpha\alpha}}$.

To assess the detectability of the cross-correlations, we computed their S/N, following the discussion of~\cite{Scelfo:2020jyw}. We organized the cross--correlation angular power spectra from different tracers and bin pairs in a vector $\mathbf{C}_\ell$ as follows:
\begin{equation}\label{eq:cvec}
\mathbf{C}_\ell = 
\begin{pmatrix}
{C}_{\ell}^{gg}(z_1, z_1), 
\dots, 
{C}_{\ell}^{gGW}(z_1, d_{L,1}), 
\dots, 
{C}_{\ell}^{GWGW}(d_{L,1}, d_{L,1}), 
\dots, 
\end{pmatrix}^{\top} \, .
\end{equation}
We associated a pair of indices $[I_1, I_2]$ to each element $I$ of this array, corresponding to the tracer and the redshift bin of the angular power spectrum in each specific entry.
With this notation, the related covariance matrix is
\begin{equation}
    \left[{\rm Cov}(\ell)\right]_{IJ} =  \tilde{C}_{\ell}^{I_1 J_1}\tilde{C}_{\ell}^{I_2 J_2} + \tilde{C}_{\ell}^{I_1 J_2}\tilde{C}_{\ell}^{I_2 J_1} \, .
\end{equation}
The diagonal entries of the covariance matrix are related to the forecasted $1\,\sigma$ error on the measured signal, (see e.g.,~\cite{Calore:2020bpd})
\begin{equation}
\begin{split}
    \delta C_\ell^{XY}(x_i,x_j) = & \Bigg\{\frac{1}{(2\ell+1)\,f_{\rm sky}}   \Big[ \Big( C_\ell^{XY}(x_i,x_j) \Big)^2  \\ 
    & + \tilde{C}_\ell^{XX}(x_i,x_i) \times \tilde{C}_\ell^{YY}(x_j,x_j)  \Big] \Bigg\}^{1/2} \, .
\end{split}
\end{equation}
We used this expression to compute the shaded bands in Fig.~\ref{fig:signal}.
The S/N for each combination of redshift and distance bins at a given multipole $\ell$ can be expressed as
\begin{equation}\label{snr_ell}
    ({\rm S/N})^2_{[I_1,I_2]}(\ell)=f_{\rm sky}(2\ell+1)\frac{\left({C}_\ell^{[I_1,I_2]}\right)^2}{\left[ \tilde{C}_\ell^{[I_1,I_1]}\tilde{C}_\ell^{[I_2,I_2]} + \left(\tilde{C}_\ell^{[I_1,I_2]}\right)^2 \right]} \, ,
\end{equation}
while the S/N of the full spectrum across all bins, taking their covariance into account, is given by
\begin{equation} \label{snr_tot}
    ({\rm S/N})^2_{\rm total}(\ell) = f_{\rm sky}\,(2\ell+1)\, \mathbf{C}_\ell^{\top} \cdot {\rm Cov}^{-1} \cdot \mathbf{C}_\ell   \, .
\end{equation}
Finally, from Eq.~\ref{snr_ell} or Eq.~\ref{snr_tot}, the cumulative S/N for $\ell<\ell_{max}$ is obtained from
\begin{equation} \label{snr_cumulative}
    ({\rm S/N})^2(\ell<\ell_{max})=\sum_{\ell'=\ell_{min}}^{\ell_{max}}({\rm S/N})^2(\ell') \, .
\end{equation}
In Fig.~\ref{fig:SNR} the color palette shows the cumulative S/N computed from Eq.~\ref{snr_ell}, while each panel shows the cumulative total S/N obtained from Eq.~\ref{snr_tot}.

\section{Surveys, error forecasts, and implementation}\label{sec:simulations}

This section provides details on the GW and galaxy catalog specifications, the error modeling, and the numerical implementation used.
\begin{figure*}[t]
  \centering
  \includegraphics[width=.4\textwidth]{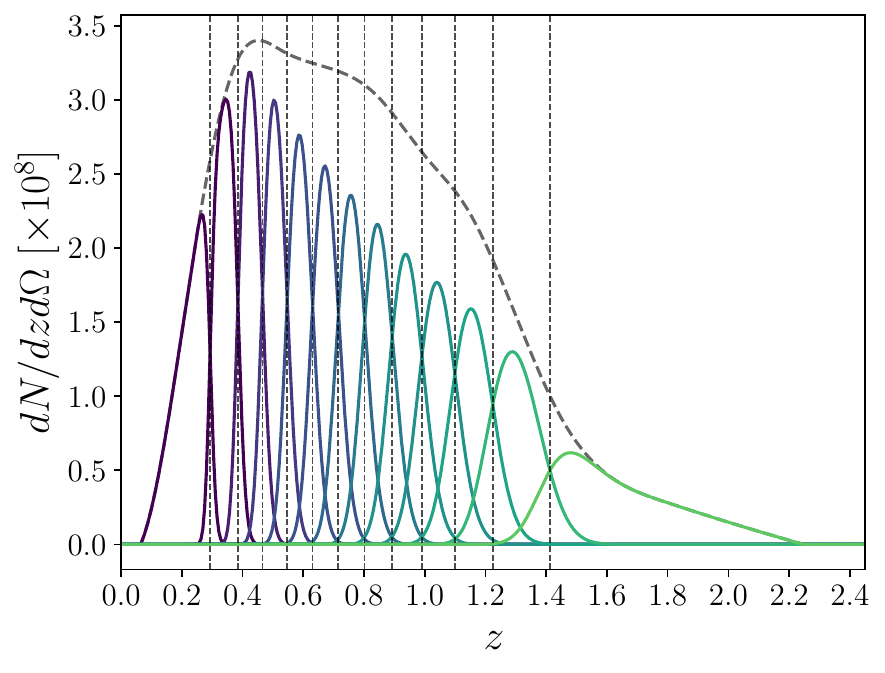}
  \hspace*{0.2cm}
  \includegraphics[width=.4\textwidth]{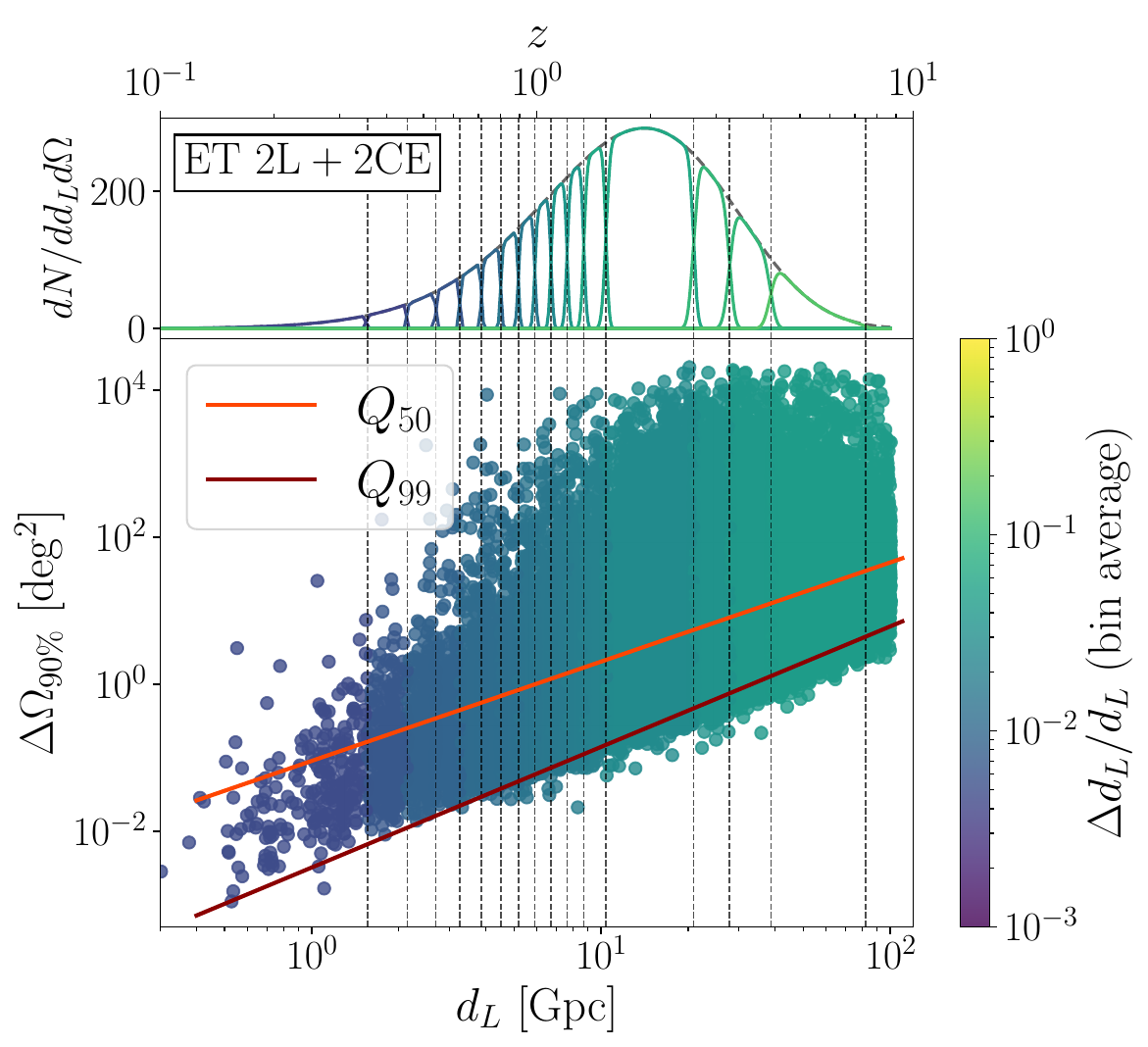}
  \caption{Left: Redshift distribution of galaxies detected by the Euclid photometric survey in 13 equally populated bins, as in the Euclid preparation study~\citep{Euclid:2021osj}. Solid colored curves represent the weighted window function in each bin (see Eq.~\ref{window}), assuming a $5\%$ relative error on the redshift. The dashed curve shows the overall event distribution. Vertical gray lines mark the bin edges.
  Right: Distribution of $90 \%$ sky localization areas of GW events detected in different distance bins.
    The color coding of the points corresponds to the relative $1\sigma$ distance uncertainty averaged over each bin.
    In the upper panel, the lines are the same as in the left panel, with the relative distance error modeled as described in the text.
    The redshift on the top axis is obtained by conversion of the distance with the fiducial cosmology.
  }
  \label{fig:gal_distr}
\end{figure*}
\subsection{Galaxy survey}\label{sec:gals}
For the galaxy survey, we considered the photometric sample of the Euclid satellite~\citep{EUCLID:2011zbd} as our fiducial scenario.
This will have a sky coverage of 15\,000 square degrees and deliver a total of ${\sim} 1.6\times10^9$ observed galaxies~\citep{EUCLID:2011zbd,Amendola:2016saw}. 
We adopted Euclid’s photometric rather than spectroscopic sample because its far larger source number reduces shot noise and its wider redshift coverage maximizes overlap with GW events. Although photometric redshifts are less precise, GW angular resolution is the dominant limitation, making the higher number of photometric sources preferable for tomographic angular power spectra.
We discuss the comparison with spectroscopic samples in Sect.~\ref{sec:specvsphot}.
For modeling the photometric survey, we relied on public results from the Euclid flagship simulation~\citep{Euclid:2021osj,Euclid:2024few}.
We refer in particular to Table I and Appendix C of~\cite{Euclid:2021rez} which provide number density, galaxy bias, and magnification bias. We interpolated the values in Table I of~\cite{Euclid:2021rez} using a cubic spline to obtain the redshift profile.
For the window function, we assumed a relative redshift uncertainty $\sigma_{\ln z}= 0.05$~\citep{Euclid:2021rez}.
As a reference case, we followed the Euclid preparation paper~\cite{Euclid:2021osj}, and adopted the standard conservative choice of 13 equally populated redshift bins.
Figure~\ref{fig:gal_distr} (left panel) shows the overall redshift distribution together with the distribution in each bin, including the effect of redshift uncertainty.
We discuss the effect of different binning choices in Sect.~\ref{sec:binning strategy}.
For both the clustering and magnification bias, we assumed the cubic polynomial fit of the Euclid flagship simulation (see Appendix C of~\cite{Euclid:2021rez}).
Values are reported in Appendix~\ref{app:popdetails}.

\subsection{Gravitational-wave survey}\label{sec:GWs}

For GW sources, we considered resolved BBH mergers.
As our default case, we considered a 3G GW detector network consisting of ET in its so-called two-L configuration (see~\cite{ET:2025xjr} for details). The two CE detectors consist of a 40 km detector and a 20 km arm detector.
We denote this configuration as ET 2L + 2CE.
We considered a five-year observing time, comparable to the duration of the Euclid survey.
We analyze the dependence on the observing time in Sect.~\ref{sec:3Gnets}, considering observing times of one and ten years and different detector configurations.
\paragraph{Catalog.} We used simulations of the expected population of astrophysical BBHs from the ET design study~\citep{Branchesi:2023mws}\footnote{Publicly available at \href{https://apps.et-gw.eu/tds/?content=3\&r=18321}{apps.et-gw.eu/tds/?content=3\&r=18321}.} (see also~\cite{Mapelli:2021syv,Mapelli:2021gyv} for details).
The catalog covers a data collection period of one year, with a total of ${\sim}1.2\times 10^5$ sources. We rescaled the amplitude of the observed source distribution to model longer observing periods.
We forecast detectability and measurement errors on distance and sky localization using the Fisher-matrix code \texttt{gwfast}~\citep{Iacovelli:2022mbg,Iacovelli:2022bbs}\footnote{The ET sensitivity curves are available at \href{https://apps.et-gw.eu/tds/?content=3\&r=18213}{apps.et-gw.eu/tds/?content=3\&r=18213}. For CE–20 km and CE–40 km we used the curves available at \href{https://dcc.cosmicexplorer.org/CE-T2000017/public}{dcc.cosmicexplorer.org/CE-T2000017/public}. These are the same sensitivity curves used in~\cite{Branchesi:2023mws,ET:2025xjr}.}. We used the \texttt{IMRPhenomXPHM} waveform model~\citep{London:2017bcn} 
and a minimum frequency of $2\,\rm Hz$ with an $85\%$ uncorrelated duty cycle in each detector in the network. 
We considered a source as detected if the optimal matched-filter S/N exceeded the threshold value of 12.
We further restricted to events with a $90\%$ credible level (CL) sky-localization area less than $4\pi$ steradians and with a relative error on the luminosity distance smaller than $100\%$\footnote{Larger localization volumes are unphysical and can appear in Fisher-matrix analyses for events for which a full Bayesian analysis would recover the prior distribution~\citep{Iacovelli:2022bbs,Dupletsa:2024gfl}.}.

\paragraph{Source distribution, distance error, and binning}
We modeled the observed source distribution with the commonly used parametrization,
\begin{equation}\label{fitGWdist}
    \frac{\dd N^{\rm GW}_{\rm obs}}{\dd d_L}=A\left(\frac{d_L}{d_{L,0}}\right)^\alpha e^{-(\nicefrac{d_L}{d_{L,0}})^\beta} \, ,
\end{equation}
where we determined the values of $\{A, d_{L,0}, \alpha, \beta\}$ from the simulations.
We report the result in Appendix~\ref{app:popdetails}, together with the total number of detected sources after applying the distance cuts and sky-localization cuts described above.
We obtained the relative error $\sigma_{\ln d_L}$ on the luminosity distance measurement in each bin, which enters the window function, as the median in the bin of $\sigma_{d_L}/d_L$, where $\sigma_{d_L}$ is the $1\sigma$ error computed from the Fisher matrix.

Figure~\ref{fig:gal_distr} (right panel) shows the weighted window function in each bin for our reference binning scheme, together with the sky-localization errors. We obtained our distance bins by converting the edges of the galaxy survey redshift bins to distance using a fiducial cosmology.
Because GW events can be detected at a larger distance than those obtained from Euclid's maximum redshift, we added three equally populated distance bins at the high-distance edge of the distribution to avoid loss of information or bias.

\paragraph{Sky localization.} As discussed in Sect.~\ref{sec:noise_and_err}, we included the impact of limited angular resolution. We imposed a cut at the maximum angular scale $\ell_{\rm max}=\pi/\Delta\Omega_{\rm min}^{1/2}$, where $\Delta\Omega_{\rm min}$ is computed in each bin from the 99th percentile of the $90\%$ CL localization error, $Q_{99}^{\Delta\Omega}(d_{L,i})$. This avoids outliers in the simulations that would yield overly optimistic predictions.
The signal is further suppressed by the exponential damping term in Eq.~\ref{shotGW}, with $\ell_{\rm damp}^2=(2\pi)^{3/2}/\Delta\Omega_{1\sigma}$. Here $\Delta\Omega_{1\sigma}$ is the 50th percentile of the $1\sigma$ localization error, $Q_{50}^{\Delta\Omega}(d_{L,i})$. We provide tables with the resulting $\ell_{\rm max}$, $\ell_{\rm damp}$, and bin edges in Appendix~\ref{app:multipoles}. Figure~\ref{fig:gal_distr} (right panel) shows the localization area of the BBH sources together with the 50th and 99th percentiles as a function of distance.
We find $\ell_{\rm damp}\sim1000$ for the closest bin ($z\in[0,0.3]$), decreasing to $\sim600$ at $z\in[0.3,0.38]$, $\sim370$ at $z\simeq0.5$, $\sim200$ at $z\simeq1$, and $\sim100$ at $z\simeq2$. 
A further cut excludes the highly nonlinear regime (Sect.~\ref{sec:summary}), which additionally tightens the multipole range at low redshift.

\paragraph{Tracer bias.} For the perturbation properties, we modeled the fiducial GW bias $b_{GW}$ as in~\cite{Peron:2023zae}, adopting the relation $b_{GW}=A_{GW}(1+z)^{\gamma}$. For the evolution and magnification bias, we followed~\cite{Zazzera:2023kjg}, which provides a third-order polynomial fit for the evolution and magnification bias for 3G detectors. We report the specific values in Appendix~\ref{app:popdetails}.

\subsection{Implementation, free parameters, and summary}\label{sec:summary}

We computed the Fisher matrix in Eq.~\ref{fisher_mat} for the auto- and cross-tomographic angular power spectra of galaxy and GW surveys as in Eq.~\ref{angular_ps}, considering the density, RSD, and lensing terms in the Limber approximation (see Appendix~\ref{app:rel_numb_count_expr}). %
We used the publicly available \texttt{python} package \texttt{COLIBRI}\footnote{\href{https://github.com/GabrieleParimbelli/COLIBRI}{github.com/GabrieleParimbelli/COLIBRI}}, which we modified to handle tracers in different spaces and to include the number-count contributions considered in this work.
The package computes the power spectrum $P(z, k)$ using \texttt{CAMB}~\citep{Lewis:1999bs}. We accounted for nonlinear scales using the \texttt{HMcode-2020} scheme~\citep{Mead:2020vgs}.
To compute the derivatives, we employed the \texttt{numdifftools} library\footnote{\href{https://github.com/pbrod/numdifftools}{github.com/pbrod/numdifftools}}. We used the central-difference method and performed a series of robustness tests to validate the stability of the numerical derivatives and the accuracy of the Fisher-matrix inversion. More specifically, for each parameter we computed derivatives over a wide range of multipoles using different step sizes, identified a stability region where the derivatives are insensitive to this choice, and selected the step size accordingly. We performed this procedure independently for each parameter. To assess the stability of the matrix inversion, we computed the quantity $\epsilon = \| F F^{-1} - I \|$, where $F$ is the Fisher matrix and $I$ is the identity matrix, and verified that it remained small in all the cases considered.

We assumed a flat $\Lambda$CDM cosmology, where the Hubble constant and the matter density, $\{H_0,\ \Omega_{\rm m}\}$, determine the distance-redshift relation. To these, we added the amplitude and spectral index of the primordial power spectrum, $\{A_s,\ n_s\}$, and the physical baryon density $\Omega_{\rm b} h^2$. We took fiducial values from Planck (2018)~\citep{Planck:2018vyg}.
We parametrized the galaxy and GW bias with one free parameter per bin for each tracer. We kept the evolution and magnification biases fixed, since they affect only the lensing term, which is subdominant, especially once the largest angular scales ($\ell\lesssim10$) are excluded.
A more general strategy would include large-scale effects and marginalize over these parameters as well~\citep{Zazzera:2024agl}. We leave this for future work.
With the choices described above, our baseline Fisher analysis contained a total of $5\ + \ N^{\rm gal.}_{\rm bins}+\ N^{\rm GW}_{\rm bins} = 8\ + \ 2\ N^{\rm gal.}_{\rm bins} $ parameters $\{\theta_\alpha\}$ (where $N^{\rm gal.}_{\rm bins}$ and $N^{\rm GW}_{\rm bins}$ are the number of bins in the galaxy and GW surveys, respectively), namely
\begin{equation}\label{params}
    \theta = \{ H_0, \Omega_{\rm m}, \Omega_{\rm b}h^2, A_s, n_s, b_{{\rm g}, 1},..., b_{{\rm g}, N^{\rm gal.}_{\rm bins}}, b_{{\rm GW}, 1},..., b_{{\rm GW}, N^{\rm GW}_{\rm bins}} \} \, .
\end{equation}
For the baseline configuration with 13 bins, this results in 34 parameters.

We restricted the multipole range to exclude highly nonlinear scales.
For bin $i$ centered at $z_i$, we adopted a maximum multipole $\ell_{\rm max}=r(z_i)\,k_{i,\rm max}$, where $r(z_i)$ is the comoving distance at the bin center and the maximum wavenumber is defined as $k_{i,\max} \equiv \pi /(2R_i)$,
where $R_i$ is defined implicitly by requiring that the root mean square (rms) linear matter-density fluctuation smoothed with a spherical top-hat of radius $R_i$ satisfies $\sigma^2(R_i,z_i)=0.25$, which is a standard choice in galaxy surveys forecasts~\citep{Amendola:2016saw}. We computed this variance from the linear matter power spectrum $P_{\rm m}(k,z)$ obtained from \textsc{camb} (as described above) as
\begin{equation}
\sigma^2(R,z)\equiv \left\langle \delta_R^2 \right\rangle
= \frac{1}{2\pi^2}\int_{0}^{\infty} \mathrm{d}k\, k^2\, P_{\rm m}(k,z)\, W^2(kR),
\end{equation}
with $W(kR)$ the Fourier-space spherical top-hat window,
\begin{equation}
W(x)=\frac{3}{x^3}\left(\sin x - x\cos x\right)
,
\end{equation}
We computed the integral numerically and solved for $R_i$ using bisection root finding.
This yields $k_{\rm max}\simeq0.12\,{\rm Mpc}^{-1}$ ($\ell_{\rm max}\simeq130$) in the lowest redshift bin and $k_{\rm max}\simeq0.21$–0.24\,${\rm Mpc}^{-1}$ ($\ell_{\rm max}\simeq800$–1000) in the highest bins.
For GW–galaxy and GW–GW correlations, $\ell_{\rm max}$ is the minimum between this nonlinearity cut and the angular scale allowed by GW localization (Sect.~\ref{sec:GWs}), with an additional exponential damping from finite sky resolution (Eq.~\ref{shotGW}). Unlike the nonlinearity cut, the localization limit decreases with redshift, reflecting poorer constraints for distant GW sources.
Overall, nonlinear scales dominate the cut at $z\lesssim0.5$, while sky resolution dominates at $z\gtrsim0.5$. A ``sweet spot'' occurs in the redshift bin $z\in[0.63,0.72]$, where $\ell_{\rm damp}=345$, maximizing the S/N of the cross–correlation (Fig.~\ref{fig:SNR}).
This multipole corresponds to the largest scale that effectively contributes to the GW-based constraints once all the discussed restrictions are accounted for\footnote{Note that these values depend strongly on the assumed GW detector network (Sect.~\ref{sec:3Gnets}).}.
\begin{figure*}[t]
  \centering
  \includegraphics[width=0.7\textwidth]{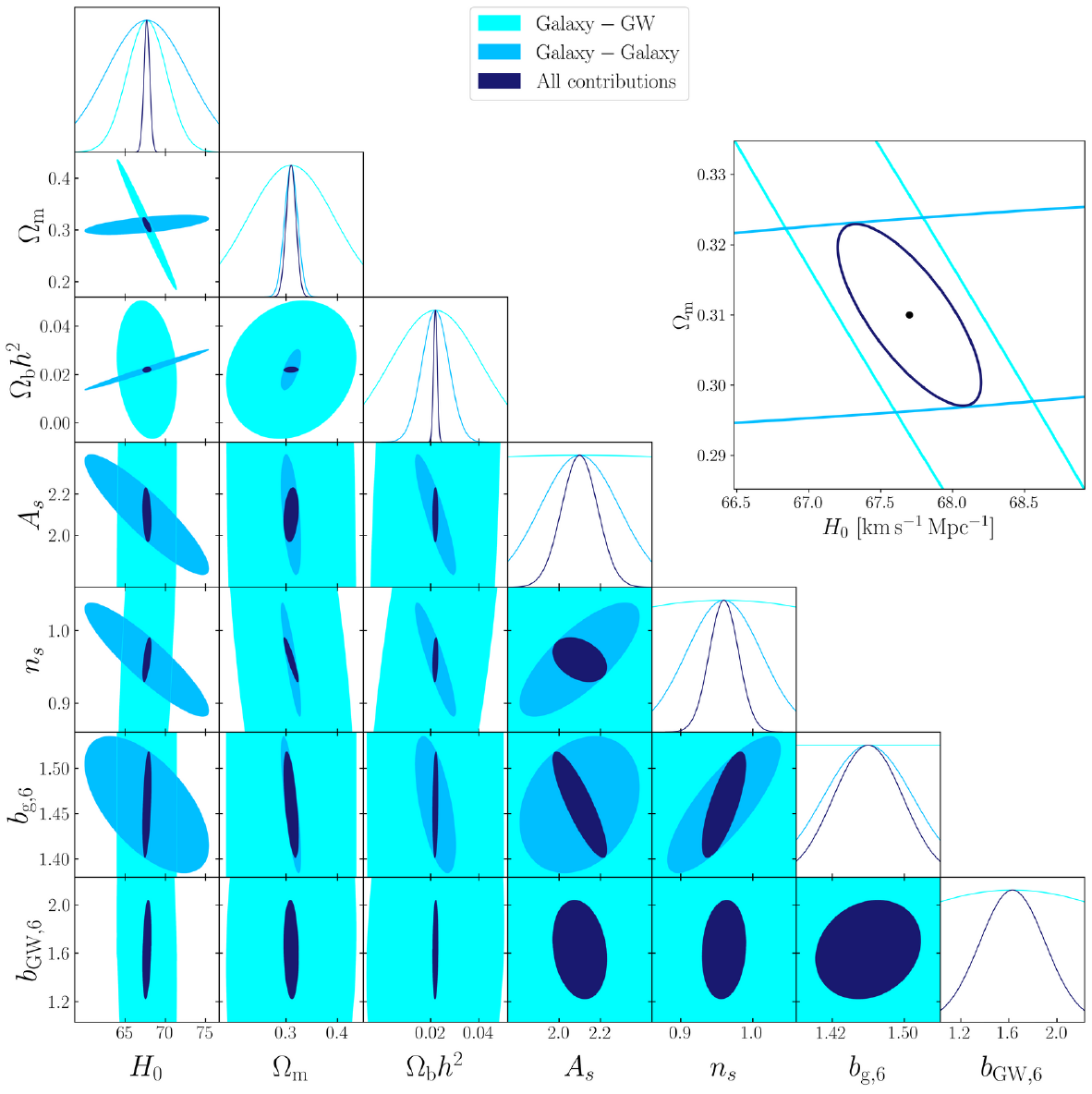}
  
  \caption{Constraints at the $1\sigma$ on cosmological and galaxy (GW) bias parameter in a selected redshift (distance) bin from the Euclid photometric survey autocorrelation (blue), the cross-correlation between Euclid and ET 2L + 2CE (cyan), and their combination including the GW-GW autocorrelation term (navy blue).
  The inset shows the $H_0-\Omega_{\rm m}$ plane with a narrower range.
  }  \label{fig:separate_contributions}
\end{figure*}

We set the minimum multipole to avoid inaccuracies of the Limber approximation. Following~\cite{Tanidis:2019teo}, which assesses the accuracy of the Limber approximation in detail, we adopted $\ell_{\rm min}=5$ for $z<0.5$, $\ell_{\rm min}=10$ for $0.5<z<0.75$, $\ell_{\rm min}=15$ for $0.75<z<1.25$, and $\ell_{\rm min}=20$ for $z>1.25$.

\section{Results and discussion}

\subsection{Euclid photometric sample combined with 3G detector networks}\label{sec:EuclidphotozETCE}

In this section we discuss results for the default configuration of this paper.
As shown in Fig.~\ref{fig:SNR}, the cross-correlation is detected with a S/N ${\sim} 21$. We also find that the GW-GW autocorrelation 
 gives a total S/N of 1.8, even though the signal in each individual multipole bin is not detectable (see Fig.~\ref{fig:signal}).

\paragraph{Constraints on the cosmic expansion}
In Figure~\ref{fig:separate_contributions}, we present the joint $1\sigma$ contours of $H_0,\, \Omega_{\rm m}$, along with other cosmological parameters and the amplitude of the galaxy and GW bias in the sixth redshift (distance) bin, chosen so to maximize the S/N for the cross-correlation\footnote{We include a free parameter quantifying the tracer bias for each bin, but showing a single one is sufficient to illustrate its effect.}.
We show contours obtained from the galaxy-galaxy and GW-galaxy correlations separately, as well as from their combination.
We report the constraints provided by each separate contribution, as well as by the combination of the probes, in Appendix~\ref{app:tables}.

The cross-correlation between GW and galaxy data (cyan) provides substantial constraining power, driven by the effect discussed in Sect.~\ref{sec:pop_pars}, namely the presence of a peak in the cross-correlation signal along the correct distance-redshift relation. This results in the typical anti-correlation between $H_0$ and $\Omega_{\rm m}$.
In particular, we find that the cross-correlation term alone provides a stronger constraint on $H_0$ than the galaxy autocorrelation.
However, our constraints for the galaxy-galaxy autocorrelation term are based on a conservative cut on the maximum multipole, selected to exclude nonlinear scales.
Regardless of this, the combination of the two probes leads to a substantial improvement in the measurement of $H_0$, due to the different correlation directions in the $H_0 - \Omega_{\rm m}$ plane.
This results in an enhancement of a factor of ${\sim} 6$ in the precision on $H_0$. Specifically, the constraint improves from ${\sim} 6\%$ when using the cross-correlation term alone (and ${\sim} 15\%$ with the galaxy-galaxy term alone) to a joint ${\sim} 1\%$ measurement (see Appendix~\ref{app:tables}).
This complementarity is illustrated in the inset of Fig.~\ref{fig:separate_contributions}, where we show the constraint in the $H_0, \Omega_{\rm m}$ plane on a narrower range.
We note that, as a consequence, the constraints on other cosmological parameters ($\Omega_{\rm b}h^2, A_s, n_s$) are also significantly improved.
We forecast an overall precision of approximately $6 \%$ for the matter density parameter $\Omega_{\rm m}$.
Overall, these constraints lead to a ${\sim}0.7 \%$ constraint on $H(z)$ at redshift $z_* \sim 0.25$, which minimizes the relative uncertainty.
We increased the sub-percent accuracy on $H_0$ and the percent-level accuracy on $\Omega_{\rm m}$ by increasing the number of bins to at least 20 (see Sect.~\ref{sec:binning strategy} below).
~\cite{Ferri:2024amc} recently forecast
 a percent-level uncertainty on $H_0$ from simulations with this technique for a similar GW detector network.
Our analysis yields a compatible precision, showing that such accuracy is achievable even when marginalizing over an extended set of nuisance parameters. For comparison, we obtain a fractional uncertainty of $\sim 0.6\%$ when fixing all nuisance parameters.

\paragraph{Astrophysical information}
\begin{figure}[t]
\centering
\includegraphics[width=0.9\columnwidth]{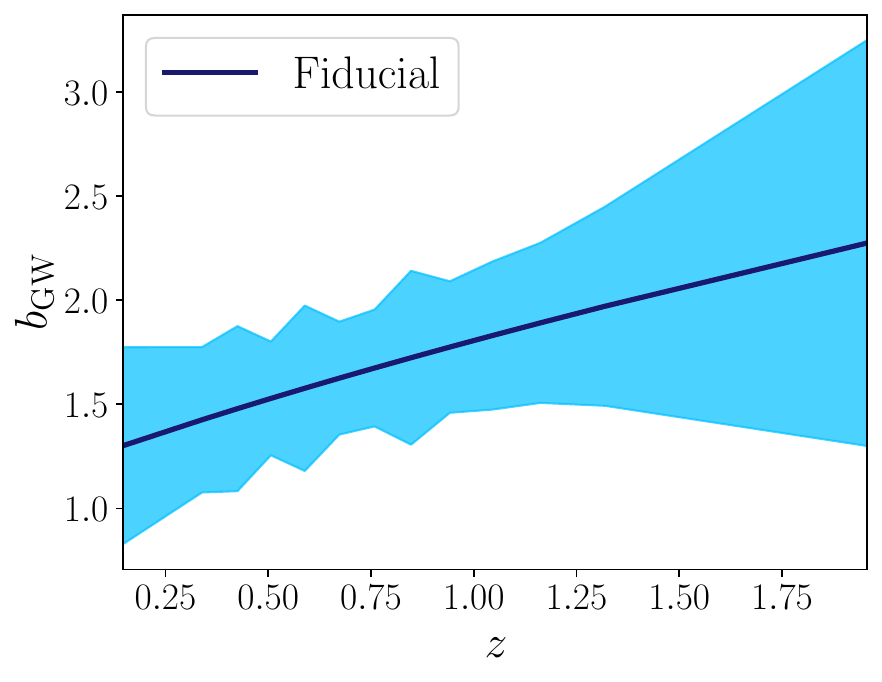}
\caption{Reconstruction of the GW clustering bias using an agnostic parametrization with one free parameter per redshift (distance) bin, using the Euclid photometric sample and five years of observations with the ET 2L + 2CE GW detector network.}
\label{fig:bias}
\end{figure}
While the main focus of this paper is the measurement of the cosmic expansion, we also present constraints on the GW clustering bias, shown in Fig.~\ref{fig:bias} as a function of redshift.
With the ET 2L + 2CE configuration we find at best a ${\sim}40\%$ constraint in $z\simeq[0.5,1]$ after five years, which improves to ${\sim}25\%$ after ten years (Appendix~\ref{app:tables}).
These bounds are less optimistic than in previous studies~\citep{Calore:2020bpd,Zazzera:2024agl}, mainly because we adopted a free bias parameter per redshift bin, as in real data analyses~\citep{Euclid:2021rez}, rather than parametric models matched to those used to obtain the fiducial values. While this widens the parameter space and weakens constraints, it is more robust against modeling systematics. Conversely, parametric assumptions can deliver tighter bounds, which may be useful for targeted astrophysical studies.
For further comparison,~\cite{Zazzera:2024agl} assume perfect knowledge of galaxy bias (but include large–scale effects beyond Limber), while~\cite{Calore:2020bpd} use a $\chi^2$ analysis fixing all other parameters. Our results are closer to~\cite{Vijaykumar:2020pzn}, who find ${\sim}20\%$ constraints from simulations with free bias parameters per bin, although using GW clustering alone in real space.

\paragraph{Dependence on the binning strategy } \label{sec:binning strategy}
\begin{figure*}[t]
  \centering
    \includegraphics[width=0.9\textwidth]{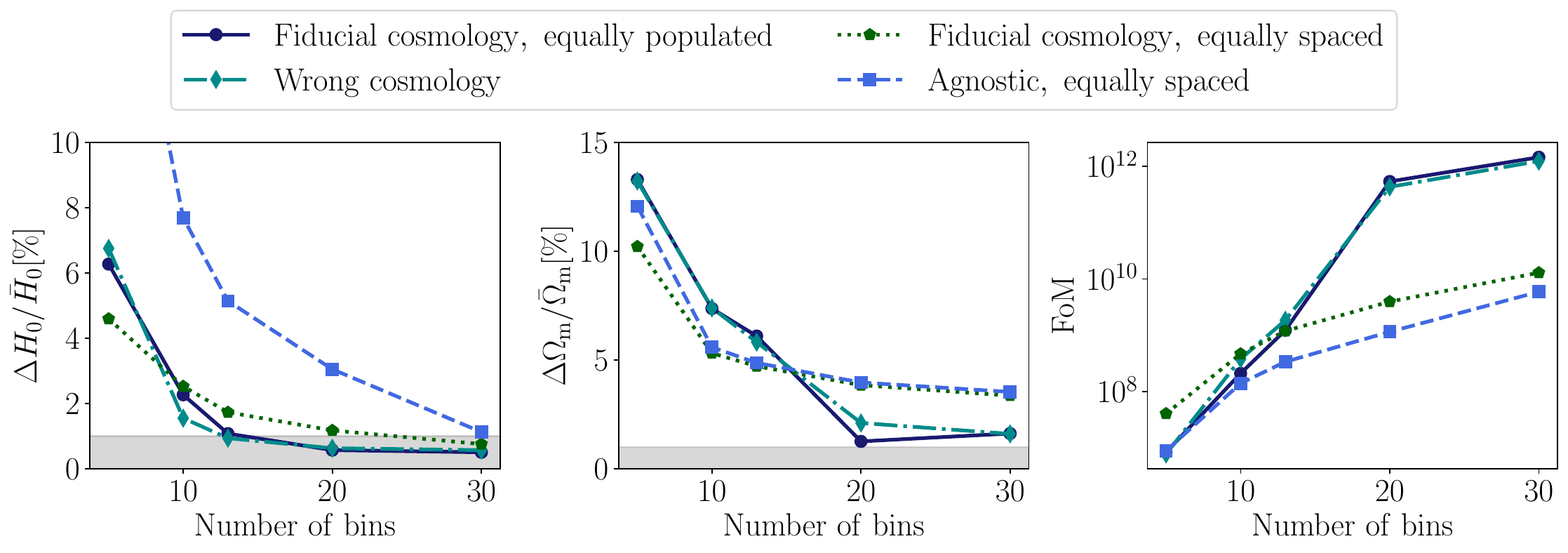}
  \caption{Relative uncertainty on $H_0$ (left) and $\Omega_{\rm m}$ (center), and ${\rm FoM}=[{\rm det} (F^{-1})]^{-1/2}$ (right), as functions of the number of bins for different binning choices described in the text, using the Euclid photometric sample and five years of observations with the ET 2L + 2CE GW detector network. Shaded horizontal bands represent $1\%$ constraints.
  A bar denotes the fiducial value.
  }
  \label{fig:nbins}
\end{figure*}
We next examine how the choice of binning, both the number of bins and their definition in redshift and distance, affects the results. We tested four strategies:

\begin{itemize}
    \item \textbf{Fiducial cosmology, equally populated (default):} sources are equally populated in redshift bins, then converted to distance with a fiducial cosmology. This would be “optimal” if the true cosmology were known.
    
    \item \textbf{Wrong cosmology:} as above, but adopting $H_0=65\,{\rm km\,s^{-1}\,Mpc^{-1}}$, $\Omega_{\rm m}=0.32$, a model excluded by joint galaxy and GW correlations.

    \item \textbf{Fiducial cosmology, equally spaced:} as the default configuration, with the difference that the bins are equally spaced in redshift.
    \item \textbf{Agnostic, equally spaced:} bins equally spaced in both redshift and distance, maximizing robustness, but yielding less information since off–diagonal covariance terms are more important.
\end{itemize}
Figure~\ref{fig:nbins} shows the relative uncertainty on $H_0$ (left) and $\Omega_{\rm m}$ (center), and the figure of merit, ${\rm FoM}=[{\rm det} (F^{-1})]^{-1/2}$~\citep{Wang:2008zh} (left), as a function of the number of bins.
Converting redshift bins to distance with a given cosmology extracts more information than agnostic binning, except at very high bin numbers where the two converge (30 bins). Using a wrong cosmology does not affect the statistical uncertainties, though a Fisher approach cannot capture systematic bias arising from such a choice.

Regarding the bin number, the uncertainty on $H_0$ converges to $\lesssim1\%$ once more than 13 bins are used: ${\sim}0.6\%$ for 20 bins and ${\sim}0.5\%$ for 30, compared to ${\sim}1.1\%$ for 13. The impact on $\Omega_{\rm m}$ is stronger: with five years and 20 bins, we obtain ${\sim}1.3\%$ precision versus ${\sim}6\%$ with 13; with one year, we obtain ${\sim}2.6\%$ on $\Omega_{\rm m}$ and ${\sim}1.3\%$ on $H_0$. This suggests that increasing beyond the 13 bins used in Euclid forecasts would substantially benefit cross–correlation with GW surveys.

Crucially, bins must be equally populated: equally spaced schemes yield much weaker improvements since constraints on $\Omega_{\rm m}$ come mainly from intermediate redshifts, and the equal number of sources keeps shot noise balanced across bins. The overall gain from 13 to 20 equally populated bins is also reflected in the FoM.
The improvement comes from better disentangling the tracer bias from $A_s$ and $n_s$. While the latter two are shared across bins, the bias is free in each bin. Thus, more bins provide independent slices where the bias can be constrained separately, reducing degeneracies with $A_s$ and $n_s$ as long as shot noise remains under control. In the limiting case of a single bin (no tomography), the bias and $A_s$ would be fully degenerate, drastically weakening constraints on other parameters.

\subsection{Spectroscopic versus photometric samples}\label{sec:specvsphot}

\begin{figure}[t]
  \centering
    \includegraphics[width=0.9\columnwidth]{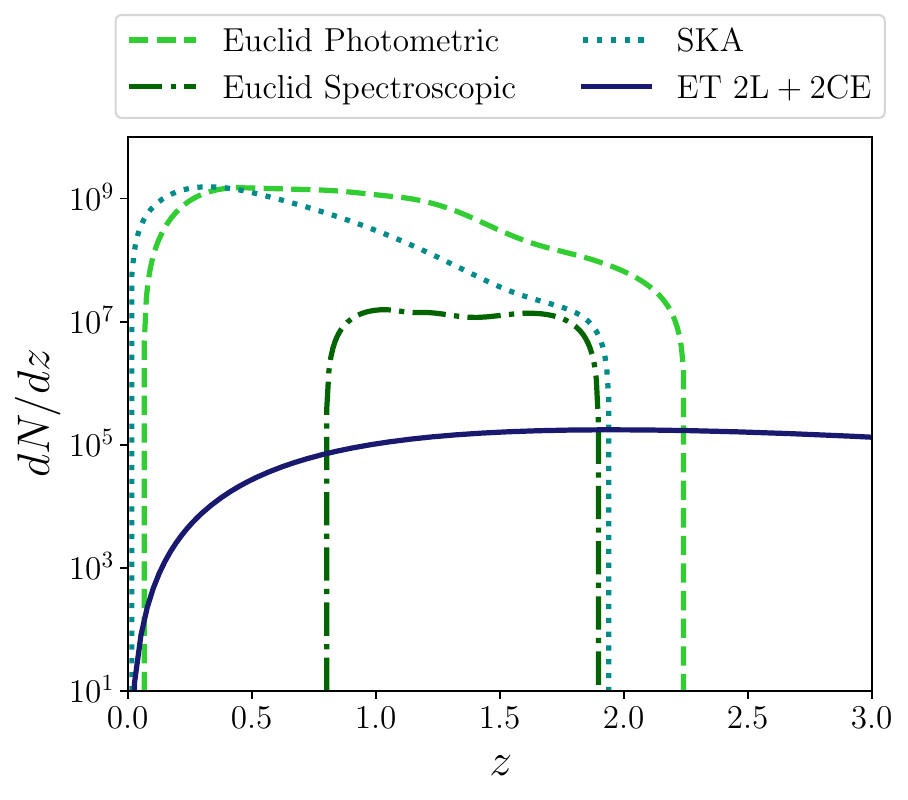}
  \caption{Redshift coverage of the different surveys considered in Sect.~\ref{sec:specvsphot}, compared with the distribution of GW sources detected over five years of ET 2L + 2CE observations, converted from distance to redshift assuming the fiducial cosmology.}
  \label{fig:zdist}
\end{figure}

In this section, we examine the impact of choosing a spectroscopic galaxy survey over a photometric one. Specifically, we compare the performance of the Euclid photometric sample to two spectroscopic surveys: the Euclid spectroscopic survey and the SKA phase II survey. In summary, we considered the following surveys:

\begin{itemize}
    \item \textbf{Euclid photometric survey (default):} this is the default choice in this paper, described in Sect.~\ref{sec:gals}.
    \item \textbf{Euclid spectroscopic survey:} 
    to model this survey, we used results derived from the Euclid flagship simulation~\citep{Euclid:2021osj,Euclid:2024few}, with fits provided in~\cite{Euclid:2023qyw}.
    In particular, we refer to their Table 2 for the expected number density, which is shown in Fig.~\ref{fig:zdist} as a dot-dashed line. This corresponds to a sample of ${\sim}1.3\times10^7$ galaxies in the redshift range $z = 0.9$ and $z = 1.8$.
    We assumed a relative redshift uncertainty of $\sigma_{\ln z} = 0.001$.
    For modeling the bias, we used the cubic polynomial fits for the linear and magnification biases provided in~\cite{Euclid:2023qyw}.
    
    \item \textbf{SKA Phase II:} to assess the relevance of a larger spectroscopic survey, we considered the optimistic scenario of the SKA Phase II instrument. This survey covers 30\,000 square degrees and is expected to detect approximately $10^9$ galaxies in the redshift range $0 \leq z \leq 2$. The survey is modeled following~\cite{Bull:2015lja}. We adopted the number density and magnification bias fitting formulae from~\cite{Maartens:2021dqy}, in particular their Eqs.~A19--A20.
\end{itemize}
\begin{figure*}[t]
  \centering
    \includegraphics[width=0.9\textwidth]{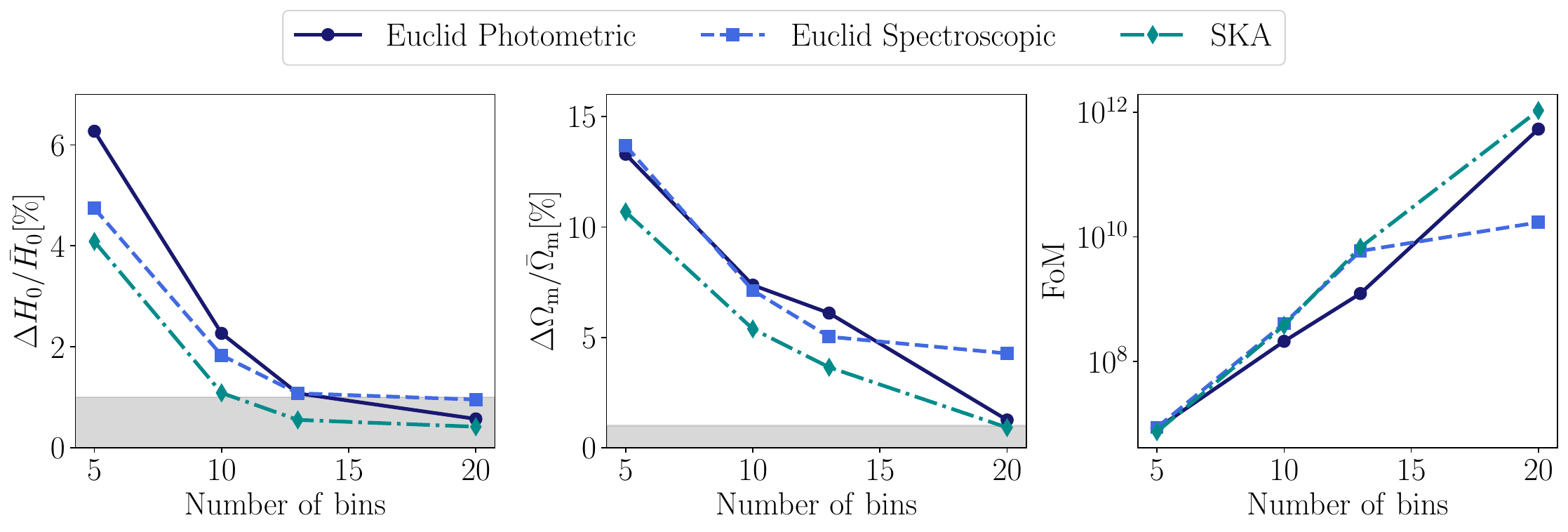}
  \caption{Relative uncertainty on $H_0$ (left) and ${\rm FoM}=[{\rm det} (F^{-1})]^{-1/2}$ (right) as a function of the number of bins for different photometric and spectroscopic surveys, using five years of observations with the ET 2L + 2CE GW detector network. Shaded horizontal bands represent $1\%$ constraints. A bar denotes the fiducial value.}
  \label{fig:specz}
\end{figure*}

Figure~\ref{fig:zdist} shows the redshift distribution of the aforementioned surveys, compared to the GW catalog considered in this work. The redshift distribution of GW sources is obtained from the distance distribution assuming a fiducial cosmology.
Figure~\ref{fig:specz} shows the relative uncertainty on $H_0$ (left) and $\Omega_{\rm m}$ (center), and ${\rm FoM}=[{\rm det} (F^{-1})]^{-1/2}$, as a function of the number of bins for the different surveys considered, combined with five years of observations of the ET 2L + 2CE configuration and the default binning scheme.

The Euclid spectroscopic sample provides constraints comparable to the Euclid photometric sample for a limited number of bins but shows saturation beyond $\sim 12$ bins.
This saturation can be attributed to the larger shot noise in the galaxy autocorrelation term, as well as to the narrower redshift range spanned by the Euclid spectroscopic survey.

In the case of SKA, we instead find increased constraining power, with a promising ${\sim} 0.4\%$ constraint on $H_0$ and a ${\sim} 0.9\%$ constraint on $\Omega_{\rm m}$ when combined with ET 2L + 2CE.
This improvement is driven by the combination of a large number of sources, comparable to the Euclid photometric sample, and a superior redshift resolution. 
However, the gain does not continue indefinitely with increasing bin number: at high resolution, the constraining power is ultimately limited by the finite sky localization of GW sources and by the nonlinearity cut applied in the analysis.
Adopting an equally spaced binning scheme might further improve the results.

We also note that in order to fully exploit the resolution of a spectroscopic survey, the correlation function may be a more promising summary statistic than the angular power
spectrum. 
With the accuracy on distance estimates provided by 3G detectors, it would be interesting to compare the performance of different summary statistics in combination with a spectroscopic survey.

\subsection{Comparison of different GW detector networks and geometries}\label{sec:3Gnets}

In this section we compare the performance of different GW detector networks combined with the Euclid photometric sample, and examine the dependence on the observation time. We considered the following configurations, which are currently under active investigation~\citep{ET:2025xjr}:

\begin{itemize}
\item $\rm{\bf{ET}}$: ET alone, either as two L-shaped detectors with 15 km arms located at the Sardinia and Meuse-Rhine candidate sites ($\rm{\bf{ET\ 2L}}$), or as a single triangular detector in Sardinia with 10 km arms ($\rm{\bf{ET\ \Delta}}$). The specific coordinates are given in~\cite{Branchesi:2023mws}.

\item $\rm{\bf{ET\ +\ 1\ CE}}$: a network of 3G detectors made of ET (in either the 2L or triangular configuration) and a single CE detector in the US with 40 km arms.
\item $\rm{\bf{ET\ +\ 2\ CE}}$: a network of 3G detectors made of ET (in either the 2L or triangular configurations, with the former being our default choice) and two CE detectors in the US with 40 km and 20 km arms.
\end{itemize}
The CE detectors locations are given in Sect.~\ref{sec:GWs}.
\paragraph{Angular resolution}
The key difference between detector configurations is their angular resolution, which we modeled as described in Sect.~\ref{sec:GWs}, with detailed results reported in Appendix~\ref{app:multipoles}. For ET alone, the median sky–localization error is substantially larger than in other setups, which strongly suppresses the signal and limits the usable multipole range. For ET $\Delta$, the damping scale is $\ell_{\rm damp}\lesssim30$ in all but the first bin, making the angular power spectrum largely uninformative. The ET 2L configuration performs better, with $\ell_{\rm damp}{\sim}160$ in the first bin but $<80$ in the others.
A more stringent source–selection cut ($\Delta\Omega_{90\%}<100\,\rm deg^2$) partially mitigates this. After removing the poorly localized events, which effectively lowers $Q_{50}$, the effective multipole range extends to $\ell_{\rm damp}{\sim}40$–80 for ET $\Delta$ and ${\sim}50$–100 for ET 2L, with a corresponding improvement in distance uncertainties\footnote{For configurations with at least one CE, this additional selection cut has negligible impact on constraining power and is therefore not applied.}. For ET–only networks, the effective multipole range is therefore determined by sky resolution rather than the nonlinearity cut. In contrast, for configurations including at least one CE detector, the transition between nonlinearity– and resolution–dominated regimes always occurs at $z\simeq0.5$–0.7.

\begin{figure}[t]
\centering
\includegraphics[width=0.9\columnwidth]{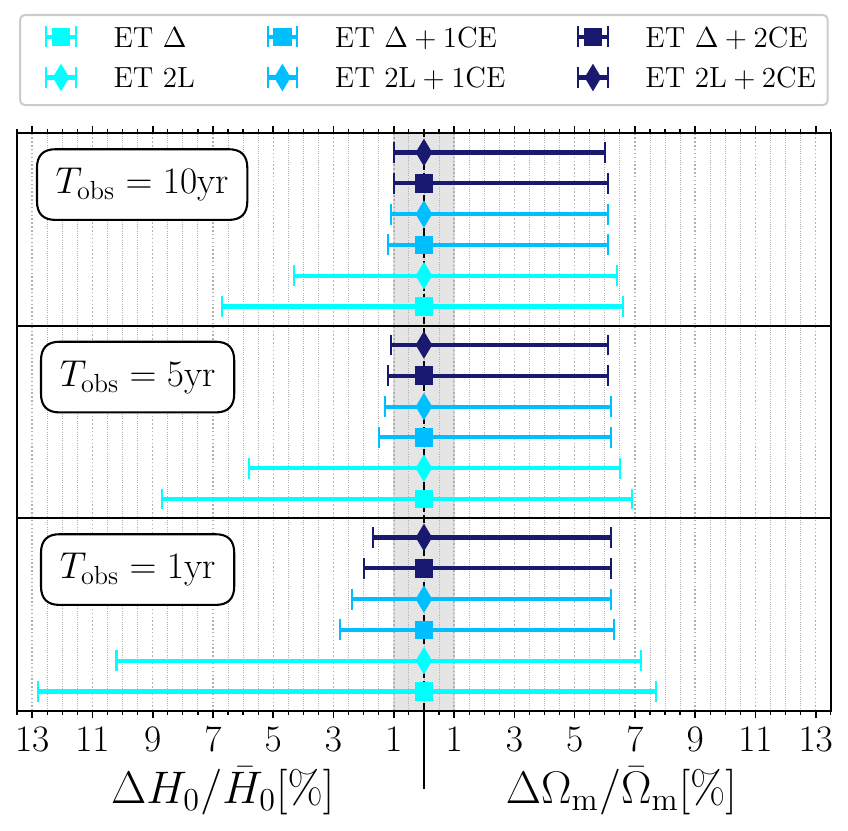}
\caption{Relative uncertainties on $H_0$ and $\Omega_{\rm m}$ forecasted for the Euclid photometric sample combined with different possible GW detector networks, for different data-taking periods.
}
    \label{fig:GW_distr_main}
\end{figure}

\paragraph{Detectability and constraints}
Across all configurations, we detect the cross–correlation after five years. The contrast between ET-only networks and those including CE is substantial: ET $\Delta$ reaches an S/N of ${\sim}1.3$ and ET 2L an S/N of ${\sim}3.7$, whereas configurations with at least one CE achieve S/N values of ${\sim}15$–21. The clustering of GW sources is not detected with S/N$>1$ for ET-only configurations, nor with ET–CE networks after only one year of observations. Detailed values are reported in Appendix~\ref{app:tables}.

Figure~\ref{fig:GW_distr_main} summarizes the forecasted marginal uncertainties on $H_0$ and $\Omega_{\rm m}$ for all configurations and observing times (one, five, and ten years). The corresponding numerical values are listed in Appendix~\ref{app:tables}. With at least one CE, percent–level precision on $H_0$ becomes achievable in five years. When a single CE is added, the combination of auto- and cross-correlations yields a factor of ${\sim}10$ improvement over either probe alone, compared to ${\sim}6$ for ET 2L + 2 CE. Adding a second CE provides smaller gains than adding the first. With one year of observations, all CE–inclusive configurations already achieve $1.7\%$ (ET 2L + 2 CE) to $3\%$ precision (ET $\Delta$ + 1 CE).

For ET alone, the cross–correlation contributes relatively little information. The resulting constraints on $H_0$ are ${\sim}70\%$ for ET $\Delta$ and ${\sim}41\%$ for ET 2L, most of the information coming from the Euclid autocorrelation. Combining with the cross–term improves the constraint by factors of ${\sim}2$ (ET $\Delta$) and ${\sim}3$ (ET 2L). The effectiveness of the cross–term clearly tracks the localization capabilities of the GW network: without triangulation from CE, the sky resolution of ET restricts the usable multipole range (Appendix~\ref{app:multipoles}), especially for ET $\Delta$. Tighter sky-localization cuts than those adopted here could be beneficial, at the cost of increased shot noise.

With ten years of observations and two CEs, the GW autocorrelation alone reaches an S/N of ${\sim}3.5$ and constrains $H_0$ to ${\sim}30\%$. In the same scenario, ET 2L alone achieves a precision of ${\sim}4\%$ on $H_0$.

\section{Conclusions}

Mapping the large-scale structure of the Universe with multiple tracers is becoming increasingly feasible.
Ongoing galaxy surveys are expected to yield high-quality datasets over the coming years, containing measurements for up to billions of redshifts. Notably, the Euclid mission~\citep{Euclid:2024yrr} is currently in its data-taking phase, and exploring the potential of combining its datasets with other probes is of great interest.
The advent of 3G GW detectors promises to significantly enhance the detection rates of resolved sources, providing precise sky localizations and distance estimates. This progress opens the possibility of using GWs as tracers of the large-scale structure (LSS).
GWs provide complementary tracers to galaxy surveys, as they probe the distance space rather than the redshift space. The conversion between distance and redshift is crucial for computing any summary statistics between the two, which causes the tomographic cross-correlation signal to peak along the distance-redshift relation (Fig.~\ref{fig:SNR}), thereby enabling constraints on cosmological parameters, as first suggested by~\cite{Oguri:2016dgk}.

In this work, we investigated the potential of cross-correlating Euclid photometric data with 3G GW catalogs following this approach. Figure~\ref{fig:separate_contributions} illustrates the complementarity of these two probes. We show that their combination provides up to a tenfold improvement in cosmological constraints relative to each probe alone.
We adopted the standard configuration from the Euclid preparation studies, e.g.,~\cite{Euclid:2021osj},~\cite{Euclid:2021rez},~\cite{Euclid:2023pyq}, which consists of 13 redshift bins. While this choice is conservative, we forecast a constraint of approximately $1\%$ on $H_0$ after five years of observations (provided that at least one CE detector is operational alongside ET), marginalized over 33 nuisance parameters. Combining with constraints on $\Omega_{\rm m}$, we forecast a $0.7\%$ constraint on the function $H(z)$ at the redshift $z_{\ast} = 0.25$, which minimizes the relative uncertainty.
If we consider a more optimistic configuration of 20 redshift bins, the constraining power improves substantially. In this scenario, we can achieve a sub-percent (${\sim} 0.6 \%$) determination of $H_0$ and a ${\sim} 1.3 \%$ constraint on $\Omega_{\rm m}$, the latter being an improvement of a factor of ${\sim} 5$ with respect to 13 bins. Such a precise measurement is only achieved with equally populated bins, which allows better suppression of shot noise.
These results demonstrate that combining Euclid photometric data with GW measurements offers competitive constraints on the cosmic expansion history based on two datasets that trace the same portion of the Universe, without relying on external priors such as those from the cosmic microwave background, supernovae, or big bang nucleosynthesis.

From an astrophysical perspective, we find that the GW source clustering alone will be detectable by ET and CE detectors after five years of observations with a maximum S/N of ${\sim} 1.8$. This would enable a measurement of the GW clustering bias, although our results are less optimistic than previous studies~\citep{Calore:2020bpd,Zazzera:2024agl} due to our agnostic assumptions on its functional form. However, they do imply that the clustering bias can be measured with a fractional error of up to ${{\sim}}40\%$ in redshift bins between $z\sim0.5-1$ after five years (see Fig.~\ref{fig:bias}), and up to ${{\sim}}25\%$ after ten years, in a model-independent way.

Moreover, we discuss the possibility of using spectroscopic galaxy catalogs, in particular the Euclid spectroscopic sample and the SKA Phase II survey. The information from the Euclid spectroscopic survey saturates around $13$ redshift bins, yielding constraints comparable to the Euclid photometric sample, while SKA Phase II could allow a sub--percent precision determination of the Hubble constant.
However, other summary statistics, such as the correlation function, may be more appropriate to fully exploit the resolution of a spectroscopic survey. This would be an interesting topic to investigate in more detail.

Finally, we show that the full potential of the tomographic cross-correlation is realized only with a network of GW detectors capable of triangulating sources and achieving good source localization.

It is important to note that the constraints derived in this study can be considered optimistic, as they are based on a Fisher matrix analysis. At the same time, we made several conservative assumptions: we excluded low multipole values consistently with the Limber approximation, although this restriction could be relaxed by including large-scale relativistic effects in the analysis; we also used a free parameter per bin for the clustering bias, without imposing priors such as continuity of the function. In addition, we restricted the analysis to linear or mildly nonlinear scales. However, these choices do not alter the conclusions regarding the complementarity between autocorrelation and cross-correlation terms.
Looking ahead, real data applications will need to address systematic uncertainties, such as the dependence on GW population properties, the impact of selection cuts, and the proper inclusion of nuisance parameters and measurement errors beyond the approximations adopted here. These issues will be crucial for fully realizing the potential of combining GWs and galaxy surveys in cosmological studies.

Finally, throughout this work we assumed a $\Lambda$CDM framework, but testing extensions is equally important.
In particular, when using GWs as probes, one must account for modified GW propagation, that is, a deviation of the GW luminosity distance from the standard relation in models with additional degrees of freedom~\citep{Belgacem:2017ihm,Belgacem:2018lbp,LISACosmologyWorkingGroup:2019mwx}.
As with $H_0$, modified GW propagation can be constrained using two–point correlations~\citep{Mukherjee:2020mha,Canas-Herrera:2021qxs,Scelfo:2022lsx,Afroz:2024joi}. This should be accompanied by scalar–sector modifications probed through galaxy surveys, so that joint GW-galaxy correlations can simultaneously probe tensor and scalar perturbations~\citep{Scelfo:2022lsx}.

Overall, two-point statistics provide a promising probe for determining the expansion history of the Universe and testing gravity at cosmological scales.

\begin{acknowledgements}
We thank Enis Belgacem, Francesco Iacovelli, Stefano Foffa, Michele Maggiore, Simone Mastrogiovanni, Isabela Matos, Niccol\'o Muttoni for discussions.
A.P, M.M., and M.S. are supported by the French government under the France 2030 investment plan, as part of the Initiative d'Excellence d'Aix-Marseille Universit\'e -- A*MIDEX AMX-22-CEI-02.
M.M. and D.G. are supported by Italian-French University (UIF/UFI) Grant No.~2025-C3-386. 
D.G. is supported by 
ERC Starting Grant No.~945155--GWmining, 
Cariplo Foundation Grant No.~2021-0555, 
MUR PRIN Grant No.~2022-Z9X4XS, 
MSCA Fellowship No.~101149270--ProtoBH,
MUR Young Researchers Grant No. SOE2024-0000125,
MUR Grant ``Progetto Dipartimenti di Eccellenza 2023-2027'' (BiCoQ),
and the ICSC National Research Centre funded by NextGenerationEU. 
This work has ET document number ET-0102A-25.
\end{acknowledgements}

\bibliographystyle{aa_edited}
\bibliography{myrefs.bib}

\appendix
\numberwithin{equation}{section}
\include{appendix}

\end{document}

%% file: appendix.tex
\section{Relativistic number counts}\label{app:rel_numb_count_expr}

In this appendix, we provide details on the calculation of the signal in the Limber approximation.

We refer to Appendix A of~\cite{Scelfo:2018sny} for the full expressions of relativistic number counts in redshift space (which follows the notation of~\cite{Bonvin:2011bg},~\cite{DiDio:2013bqa}), and to Appendix B of~\cite{Fonseca:2023uay} for luminosity distance space. Note that~\cite{Fonseca:2023uay} report directly the transfer functions, see~\ref{eq:source_num_den}, using conformal time rather than redshift as integration variable. The expressions for the perturbations in luminosity distance space as functions of redshift can be read from those with the appropriate substitution of the jacobian from distance to conformal time with the one from distance to redshift. 
Explicitly, the density, RSD/LSD, and lensing contributions 
that we use in this work, read
    \begin{eqnarray}
       \Delta_\ell^{X, {\rm den}}(k, z) & = & b_{X}(z)\, \mathcal{T}_{\rm m}(z,k)\, j_{\ell}(kr(z))  \, , \label{density}\\
       \Delta_\ell^{X, {\rm len}}(k, z)  & = & \frac{\ell(\ell+1)}{r(z)}\, \int_0^{r(z)} \dd r \, j_{\ell}(kr)\,  \\ & \times & A_L^X\big(r, r(z) \big)\, \mathcal{T}_{\Phi+\Psi}(z,k) \, \label{lensing}\\
        \Delta_\ell^{X, {\rm RSD/LSD}}(k, z)  & = & A^{X}_{\rm LSD/RSD}(z) \,  k\mathcal{T}_V(z, k) \, j^{''}_{\ell}(kr(z))\, \label{RSD}
     \end{eqnarray} \,
    where $r(z)=c\int_0^z dz'\,H^{-1}(z')$ is the comoving distance. The lensing and RSD/LSD coefficients $A_L^X$, $A^X_{\rm RSD/LSD}$ are given for the two tracers by
\begin{align} 
    & A_L^{\rm gal}(r, \bar{r})=\frac{1}{2}(5s^{\rm g}-2)\frac{\bar{r}-r}{r} \, , \\
    & A_L^{\rm GW}(r, \bar{r}) =  \frac{1}{2}\left[\left(\frac{\bar{r}-r}{r}\right)(\beta-2)+\frac{1}{1+\bar{r}\mathcal{H}}\right]\, , \\ \label{len_gal}
    & A_{\rm RSD}^{\rm gal} = \frac{1}{\mathcal{H}} \, , \\
    & A_{\rm LSD}^{\rm GW} =  \frac{2 \, \gamma}{\mathcal{H}} \, ,  \label{rsd_a}
\end{align}
with
\begin{align} \label{bias_GW}
    & \gamma=\bar{r}\mathcal{H}/(1+\bar{r}\mathcal{H}) \, , \\
    &  \beta=5s^{\rm GW}-1+\gamma\left[\frac{2}{\bar{r}\mathcal{H}}+\gamma\left(\frac{\dot{H}}{H^2}+1-\frac{1}{\bar{r}\mathcal{H}}\right)-1-b_e^{\rm GW}\right] \, ,
\end{align}
where $\mathcal{H}(z)=H(z)/(1+z)$, $b_e^{\rm GW}$ and $s^{\rm GW}$ are respectively the evolution and magnification bias for GWs, and $s^{\rm g}$ is the magnification bias for galaxies. Finally, we denoted by $\mathcal{T}_{\Phi+\Psi}(z,k)$ and $\mathcal{T}_{\rm m}(z,k)$ the transfer functions for the potentials $\Phi$, $\Psi$ and for matter respectively. Those are related to the initial curvature perturbation by 

\begin{align}
&(\Phi+\Psi)(k,z) = \mathcal{T}_{\Phi+\Psi}(z,k) \Psi_{\rm in}(k)\, , \\
&\delta_{\rm m}(k,z) = \mathcal{T}_{\rm m}(z,k) \Psi_{\rm in}(k)
\end{align}
where $\Psi_{\rm in}(k)$ has power spectrum $k^3 \langle \Psi_{\rm in}(\vb{k}) \Psi^*_{\rm in}(\vb{k'}) \rangle = (2\pi)^3 \delta(\vb{k}-\vb{k'}) \mathcal{P}(k)$ with $\mathcal{P}(k)=k^3 P(k) = A_s (k/k_*)^{n_s-1}$. 

    We use the above expressions to compute the power spectra in Eq.~\ref{angular_ps} using the Limber  approximation~\citep{Limber:1954zz}. In this case, one can evaluate integrals of the form~\ref{angular_ps} by making use of the asymptotic formula~\citep{LoVerde:2008re}:
\begin{equation}
    \frac{2}{\pi}\int \dd k \,  k^2\, j_l(kr)\, j_l(kr')\simeq\frac{1}{r^2}\delta(r-r') +\mathcal{O}\left(\ell+1/2\right)^{-2}\, ,
\end{equation} 
 and evaluating the power spectrum at the wavenumber
which projects on the angular scale $\ell$, i.e.:
\begin{equation}
    P(k, z)=P\left[k=\frac{\ell+1/2}{r(z)}, z\right] \, .
\end{equation}
In GR one has $\Phi=\Psi$, and the potentials are related to the matter transfer function via the Poisson equation:
\begin{equation}
    \mathcal{T}_{\Phi+\Psi} = -\frac{3 \, H_0^2 \, \Omega_{\rm m}}{k^2}\, (1+z)\, \mathcal{T_{\rm m}} \, .
\end{equation}

For the density and lensing terms, we obtain (see also~\cite{Euclid:2021rez})
\begin{equation} \label{C_l _limber_denden}
    C_\ell^{{\rm den}X, {\rm{den}} Y}(x_i,x_j) = \int_0^\infty \frac{c\,dz}{H(z)\,r^2(z)}\, \tilde{W}^{X}(z, x_i) \tilde{W}^{Y}(z, x_j)\, P\left(\frac{\ell+1/2}{r(z)}, z\right) \, , 
\end{equation}
with $\tilde{W}^X(z, x_j) =  J_X(z)\, b_X(z)\, w^{X}(z, x_i)\, H(z) /c$. Then,

\begin{equation} \label{C_l _limber_ll}
\begin{split}
    C_\ell^{{\rm len}X, {\rm{len}} Y}(x_i,x_j) = & \, \frac{\ell^2(\ell+1)^2}{\left(\ell+1/2\right)^4}\, \left(\frac{3 \, H_0^2 \, \Omega_{\rm m}}{c^2}\right)^2\, \int_0^\infty  \frac{\dd z_1}{r(z_1)}  J_X(z_1) \\
    & \times  \, \, w^{X}(z_1, x_i) \, \int_0^{z_1} \frac{\dd z_2}{r(z_2)}\, J_Y(z_2) \, w^{Y}(z_2, x_j) \\ 
    & \times \int_0^{{\rm min}(z_1, z_2)} \frac{c\,\dd x}{H(x)}\, r(x)^2\, A_L^{X}\Big(r(x), r(z_1)\Big) \,  \\
    & \times A_L^{Y}\Big( r(x), r(z_2)\Big) \, (1+x)^2\, P\left(\frac{\ell+1/2}{r(x)}, x\right) \, , 
    \end{split}
\end{equation}
and
\begin{equation} \label{C_l _limber_dl}
\begin{split}
    C_\ell^{{\rm len}X, {\rm{den}} Y}(x_i,x_j) = & -\frac{\ell(\ell+1)}{\left(\ell+1/2\right)^2}\, \frac{3 \, H_0^2 \, \Omega_{\rm m}}{c^2} \, \int_0^\infty  \frac{\dd z_1}{r(z_1)} \, J_X(z_1)  \\
    & \times \,w^{X}(z_1, x_i)  \, \int_0^{z_1} \dd z_2\, J_Y(z_2) \, w^{Y}(z_2, x_j) \, b_{X}(z_2)  \\ 
    &\times  \, (1+z_2) \, A_L^{Y}\Big(r(z_2), r(z_1)\Big) \, P\left(\frac{\ell+1/2}{r(z_2)}, z_2\right) \, , 
    \end{split}
\end{equation}
In the above equations, $J_{X}(z)$ denotes a jacobian factor that depends on whether the tracer $X$ is in redshift (galaxies) or distance (GWs) space, with values $J_{\rm g}(z) = 1$, $J_{\rm GW}(z) = \dd d_L/\dd z$.

For the RSD/LSD term, the transfer function of the velocity is related to the matter one by the continuity equation
\begin{equation}
    k \mathcal{T}_V(z, k) = - \frac{H(z)}{1+z} \, f(z) \, \mathcal{T}_m(z) \, ,
\end{equation}
with $f(z) \equiv \dd \log \mathcal{T}_m(z) / \dd \log a$ being the growth function.
One can then use a recurrence relation for the spherical Bessel functions to express $j_{\ell}''$ as a combination of $j_{\ell}$ which receives contributions from $\ell = -2,\ -1,\ 0,\ 1,\ 2$~\citep{Tanidis:2019teo,Euclid:2023pyq}.
~\cite{Tanidis:2019teo} showed that, for tracers in redshift space, the result can be expressed in the form of Eq.~\ref{C_l _limber_denden}, with a suitable definition of the window function, namely

\begin{equation}
\begin{split}
    \tilde{W}_{\ell, \rm RSD}^X( z, z_j ) = &\sum_{k=-1}^{1} \, L_{k}(\ell) \, w^X\left[ \frac{2\ell+1+4 k}{2\ell+1} r(z) , z_j \right] \,\\
    &\times  f\left[ \frac{2\ell+1+4 k}{2\ell+1} r(z) \right] \, ,
    \end{split}
\end{equation}
with 
\begin{equation}
    \begin{split}
        L_{0}(\ell) & = \frac{2\ell^2+2\ell-1}{(2\ell-1)(2\ell+3)} \, ,\\ 
        L_{1}(\ell) & = \frac{(\ell+2)(\ell+1)}{(2\ell+3)\sqrt{(2\ell+1)(2\ell+5)}} \, ,\\
        L_{-1}(\ell) & = - \frac{(\ell+2)(\ell+1)}{(2\ell-1)\sqrt{(2\ell-3)(2\ell+1)}} \, .\\
    \end{split}
\end{equation}
For tracers in distance space, the only difference is that the LSD term has a different prefactor, cf. Eq.~\ref{rsd_a}, and carries an extra jacobian factor from distance to redshift space. As a result, the expression for LSD is 
\begin{equation}
\begin{split}
        \tilde{W}_{\ell, \rm LSD}^X( z, d_{L,j} ) = & \sum_{k=-1}^{1} \, L_{k}(\ell) \, w^X \\
        & \times \left[ \frac{2\ell+1+4 k}{2\ell+1} r(z) , d_{L,j} \right] \, \left( 2 f\, \gamma \, \left.\frac{\dd d_L}{\dd z}  \right) \right|_{\frac{2\ell+1+4 k}{2\ell+1} r(z)} \, ,
        \end{split}
\end{equation} 
with $\gamma$ given in Eq.~\ref{bias_GW}.
Finally, the growth rate is given in $\Lambda$CDM by $f(z)=\Omega_{\rm m}(z)^{6/11}$~\citep{Wang:1998gt}.
The expressions for the RSD/LSD terms and their cross--terms with density and lensing contributions are then obtained by replacing the appropriate window functions in Eqs.~\ref{C_l _limber_denden}-\ref{C_l _limber_dl}.

\section{Details on population modeling}\label{app:popdetails}

In this appendix we provide details on the modeling of the galaxy and gravitational--wave catalogs and the respective tracer biases.

\paragraph{Euclid photometric catalog bias.} The linear and magnification bias for the Euclid photometric survey are:
\begin{equation}
\begin{split}
    & b_{\rm g}(z)  =  b_{\rm g, 0} + b_{\rm g, 1} z + b_{\rm g, 2} z^2 + b_{\rm g, 3} z^3 \, , \\ 
    & b_{\rm g, 0} = 0.5125, \, b_{\rm g, 1} = 1.377, \, b_{\rm g, 2} = 0.222, \, b_{\rm g, 3} = -0.249 \, ,\\
    & s_{\rm g}(z)  = s_{\rm g, 0} + s_{\rm g, 1} z + s_{\rm g, 2} z^2 + s_{\rm g, 3} z^3 \,, \\
    & s_{\rm g, 0} = 0.0842, \, s_{\rm g, 1} = 0.0532, \, s_{\rm g, 2} = 0.298, \, s_{\rm g, 3} = -0.0113 \,.
\end{split}
\end{equation}
\paragraph{Observed distributions of GW sources.} The fits to the observed distribution of GW sources (see Eq.~\ref{fitGWdist}) is provided in Tab.~\ref{tab:fit_parameters}.

\begin{table*}[t]
\caption{Parametric fit to the observed distribution of GW sources.}
\centering
\begin{NiceTabular}{c||c|c c c c}
\multirow{2}{*}{Detector network} & \multirow{2}{*}{$N^{\rm GW}_{\rm obs}$} & \multicolumn{4}{c}{Best--fit parameters}\\
& & $A\ \rm [Gpc^{-1}]$ & $d_{L,_0}\ \rm [Gpc]$ & $\alpha$ & $\beta$ \\
\hline
\hline
ET $\Delta$\ 10\ km & $5.3\times 10^4$ & 330.897 & 6.352 & 1.697 & 0.922\\
$\rm{ET\ 2L\ 45^{\circ}\ 15\ km}$ & $6.1\times 10^4$ & 155.232 & 3.466 & 2.152 & 0.765\\
ET $\Delta$\ 10\ km (localization cut) & $1.05\times 10^4$ & 99.0 & 6.89 & 1.25 & 0.97\\
$\rm{ET\ 2L\ 45^{\circ}\ 15\ km\ (localization\ cut)}$ & $1.7\times 10^4$ & 61.34 & 1.97 & 1.93 & 0.7\\
ET $\Delta$\ 10\ km + 1 CE & $1.0\times 10^5$ & 69.695 & 1.79 & 2.539 & 0.658\\
$\rm{ET\ 2L\ 45^{\circ}\ 15\ km\ +\ 1\ CE}$ & $1.03\times10^5$ & 626.25 & 1.533 & 2.619 & 0.658\\
ET $\Delta$\ 10\ km + 2 CE & $1.1\times 10^5$ & 40.143 & 1.364 & 2.693 & 0.625\\
$\rm{ET\ 2L\ 45^{\circ}\ 15\ km\ +\ 2\ CE}$ & $1.12\times10^5$ & 412.11 & 1.244 & 2.729 & 0.614 \\
\end{NiceTabular}%
\tablefoot{Coefficients of the parametric fit Eq.~\ref{fitGWdist} a function of the luminosity distance, for the different observatories considered in this work, and one year of observation.}
\label{tab:fit_parameters}
\end{table*}

\paragraph{GW bias.} The perturbation properties of GW sources are modeled as follows.

\begin{itemize}
\item Clustering bias $b_{\rm  GW}$:
    \begin{equation} \label{bias_gw}
        b_{\rm  GW}=A_{\rm  GW}(1+z)^{\gamma} \, ,
    \end{equation} 
    with fiducial values $A_{\rm  GW}= 1.20$, $\gamma=0.59$~\citep{Peron:2023zae}. 
\item Evolution and magnification bias~\citep{Zazzera:2023kjg}:
\begin{equation}\label{eq:GWbiasfit}
\begin{split}
     b_{{\rm e}, {\rm GW}}(z) & =  b_{{\rm e}, {\rm GW,0}} + b_{{\rm e}, {\rm GW,1}} z + b_{{\rm e}, {\rm GW,2}} z^2 + b_{{\rm e}, {\rm GW,3}} z^3  \, ,\\
    s_{\rm GW}(z) & =  s_{\rm GW, 0} + s_{\rm GW, 1} z + s_{\rm GW, 2} z^2 + s_{\rm GW, 3} z^3 \, .
\end{split}
\end{equation}
The coefficients depend on the configuration, and are reported in Tab.~\ref{tab:bias_GW} for an S/N threshold of 12 for GW sources.
\end{itemize}

\begin{table*}[t]
\caption{Evolution and magnification bias fits.}
\centering
\resizebox{\textwidth}{!}{
\begin{NiceTabular}{c||c c c c|c c c c}
\multirow{2}{*}{Detector network} & \multicolumn{4}{c}{$b_{{\rm e}, {\rm GW}}$} & \multicolumn{4}{c}{ $s_{\rm GW} \times 10^{3}$}\\
&  $b_{{\rm e}, {\rm GW,0}}$ & $b_{{\rm e}, {\rm GW,1}}$ & $b_{{\rm e}, {\rm GW,2}}$ & $b_{{\rm e}, {\rm GW,3}}$ & $s_{\rm GW,0}$ & $s_{\rm GW,1}$ & $s_{\rm GW,2}$ & $s_{\rm GW,3}$ \\
\hline
\hline
ET alone & -1.45 & -1.39 & 1.98 & $-3.63 \times 10^{-1}$ & $-8.3$ & $45.4 $ & $13.6$ & $-2.04$\\
ET + CE & -1.45 & -1.39 & 1.98 & $-3.63 \times 10^{-1}$ & $-5.59$ & $29.2$ & $3.44$ & $2.58$ \\
\end{NiceTabular}
}
\tablefoot{Coefficients of the evolution and magnification bias fits in Eq.~\ref{eq:GWbiasfit} for GW sources, from~\cite{Zazzera:2023kjg}}
\label{tab:bias_GW}
\end{table*}

\section{Effect of limited sky resolution}\label{app:multipoles}

We detail here the impact of GW sky–localization accuracy on the estimation of the angular multipoles of galaxies.

We start from the probability distribution  $p$ for the reconstructed position of a GW event, assumed axis–symmetric around the true source and peaked at that location. We model it as a Gaussian in $\mu=\cos\theta$ (where $\theta$ is the polar angle):
\begin{equation}
p(\theta, \phi) = \frac{2}{(2\pi)^{3/2}\sigma_\mu}e^{-(\cos\theta-1)^2/(2\sigma_\mu^2)},
\end{equation}
with flat distribution in the azimuthal angle $\phi$. Normalization ensures that the integral over $\theta$ and $\phi$ is unitary, assuming $5\sigma_\mu<2$. The parameter $\sigma_\mu$ is related to the $1\sigma$ solid angle $\Delta\Omega_{1\sigma}=2\pi\sigma_\mu$.

We stress that the modeling of the sky–localization kernel necessarily involves an approximation. In GW observations, one reconstructs a posterior distribution over directions from the strain data, mainly driven by triangulation. Within a Fisher-matrix treatment, this posterior is approximated as Gaussian in a chosen angular variable, which is not unique. One may equivalently adopt Gaussianity in $\theta$ or in $\mu=\cos\theta$. Here we adopt a Gaussian in $\mu$, which is natural given the uniform distribution of sources on the sphere and allows for an analytic treatment of the angular integrals. We emphasize that alternative prescriptions would be equally legitimate: all such constructions rely on approximations and correspond to different choices of which variable is treated as Gaussian. Our choice aims to provide a transparent, consistent, and reproducible derivation of the damping effect.

The observed field $\tilde\delta(\theta,\phi)$ in the sky is the convolution of the true field with $p$, rotated to align with $(\theta,\phi)$:
\begin{equation}
\tilde\delta(\theta, \phi) = \int \delta(\theta',\phi'|\theta,\phi)\,p(\theta',\phi')\,d\Omega.
\end{equation}
Expanding both $\delta$ and $p$ in spherical harmonics and using Wigner–D matrices yields
\begin{equation}
\tilde \delta(\theta, \phi) = \sum_{\ell m m_1} (-1)^m \delta_{\ell m_1} p_{\ell m}[D_{m_1 -m}^{(\ell)}]^* .
\end{equation}
The axial symmetry of $p$ implies $p_{\ell m}=p_\ell\delta_{m0}$, giving
\begin{equation}
\tilde\delta_{\ell m}=\tilde p_\ell \delta_{\ell m},\qquad
\tilde p_\ell = \sqrt{\frac{4\pi}{2\ell+1}}p_\ell .
\end{equation}
Thus $\tilde p_\ell$ acts as a damping function, which explicit form is given by
\begin{equation}
\tilde p_\ell = \frac{2}{\sqrt{2\pi}\sigma_\mu}\int_{-1}^1 L_\ell(\mu)e^{-(\mu-1)^2/(2\sigma_\mu^2)}d\mu ,
\end{equation}
where $L_\ell$ are Legendre polynomials. Using their series expansion gives
\begin{equation}\label{explicit}
\tilde p_\ell = \frac{2}{\sqrt{2\pi}\sigma_\mu}\sum_{k=0}^\ell \frac{(\ell+k)!}{(k!)^2(\ell-k)!}\frac{(-1)^k}{2^k}\sigma_\mu^{k+1}I_k,
\end{equation}
with $I_0=\sqrt{\pi/2}$, $I_1=1$, and $I_k=I_0(k-1)!!$ for even $k$, $I_k=I_1(k-1)!!$ for odd $k$. For $\sigma_\mu\ll1$, higher–order terms are suppressed, and at low $\ell$:
\begin{equation}
\tilde p_\ell\simeq 1-\ell(\ell+1)\frac{\sigma_\mu}{\sqrt{2\pi}}
\simeq e^{-\ell(\ell+1)/\ell_{\rm damp}^2},
\end{equation}
with
\begin{equation}
\ell_{\rm damp}^2=\frac{\sqrt{2\pi}}{\sigma_\mu}=\frac{(2\pi)^{3/2}}{\Delta\Omega_{1\sigma}}.
\end{equation}
The above equation shows that the damping multipole $\ell_{\rm damp}$ is related to the $1$-sigma error region in the localization of GW events.

\begin{figure*}[t]
\centering
    \includegraphics[width=0.9\textwidth]{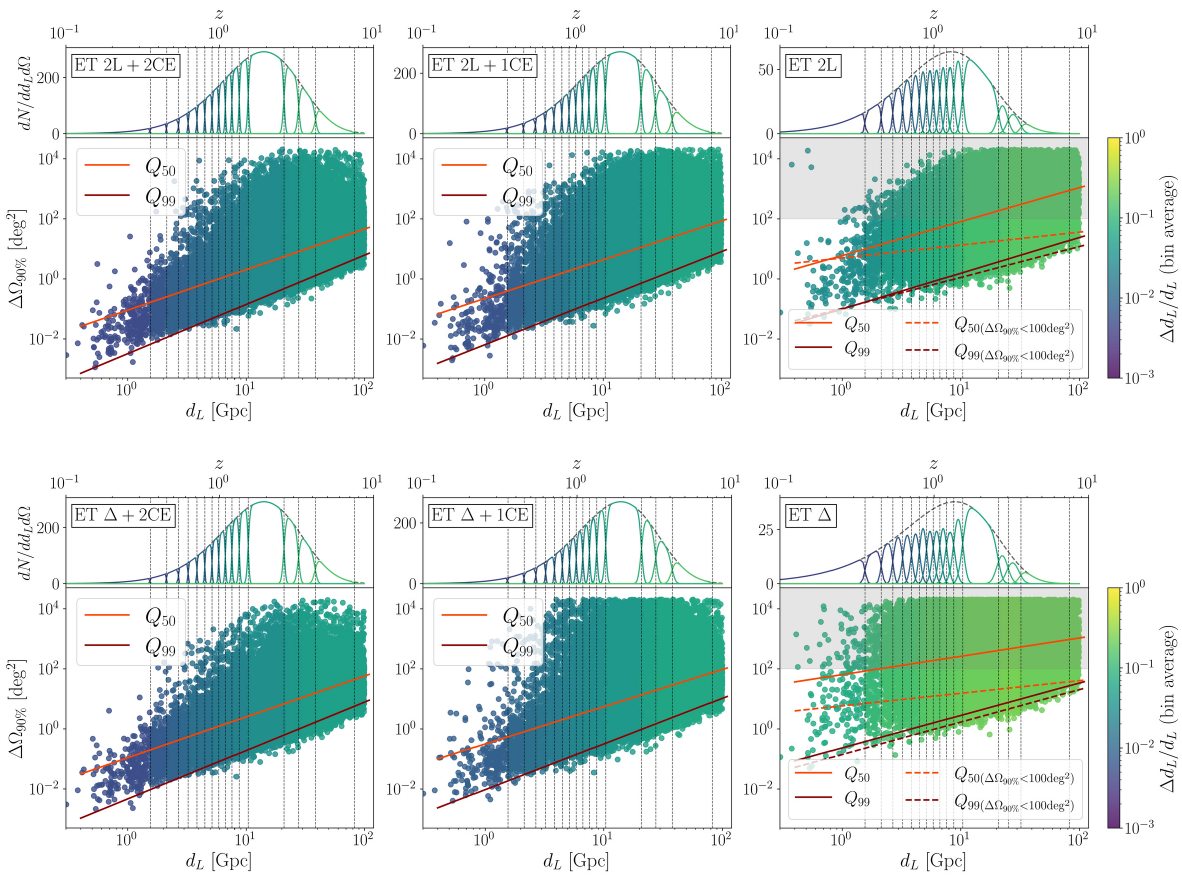}
        \caption{As in the right panel of  Fig.~\ref{fig:gal_distr}, for different GW detector networks considered in this work. Shaded areas in the left column correspond to sources with $90 \%$ sky localization $\Delta\Omega_{90 \%}<100\, \rm deg^2$, which we exclude from the analysis.}
    \label{fig:GW_distr}
\end{figure*}

Tables~\ref{tab:valuesD}–\ref{tab:values2L} list the multipole and wavenumber cuts adopted in this work, including $\ell_{\rm max}^{\rm NL}$ from nonlinearity, $\ell_{\rm max}^{\rm LOC}$ from GW angular resolution, and $\ell_{\rm damp}$ from finite sky localization.

Figure~\ref{fig:GW_distr} shows the localization areas, together with their 50th and 99th percentiles as function of distance, the source distributions and bins, for the different GW detector configurations considered in the paper.

\begin{table*}[t]
    \caption{Bin edges for ET $\Delta$.}
    \centering
    \resizebox{\textwidth}{!}{
    \begin{NiceTabular}{c||cc|cc|cc|cc|cc}
        & & & \multicolumn{2}{c}{ET $\Delta$} & \multicolumn{2}{c}{ET $\Delta$ (loc. cut)} & \multicolumn{2}{c}{ET $\Delta$ + 1 CE} & \multicolumn{2}{c}{ET $\Delta$ + 2CE}\\
        \textbf{Bin Edges} & \textbf{$k_\text{max}^\text{NL}$} & \textbf{$\ell_\text{max}^\text{NL}$} & \textbf{$\ell_\text{damp}$} & \textbf{$\ell_\text{max}^\text{LOC}$} & \textbf{$\ell_\text{damp}$} & \textbf{$\ell_\text{max}^\text{LOC}$} & \textbf{$\ell_\text{damp}$} & \textbf{$\ell_\text{max}^\text{LOC}$} & \textbf{$\ell_\text{damp}$} & \textbf{$\ell_\text{max}^\text{LOC}$} \\
        \hline
        \hline
        $0 - 0.3$ & 0.117 & 130 & 50 & 634 & 108 & 726 & 677 & $>1000$ & 1040 & $>1000$ \\
        $0.3 - 0.38$ & 0.122 & 171 & 31 & 294 & 85 & 304 & 401 & $>1000$ & 573 & $>1000$ \\
        $0.38 - 0.47$ & 0.128 & 214 & 20 & 259 & 65 & 305 & 306 & 950 & 416 & $>1000$\\
        $0.47 - 0.55$ & 0.135 & 263 & 19 & 223 & 74 & 241 & 286 & 740 & 395 & 897 \\
        $0.55 - 0.63$ & 0.142 & 315 & 22 & 171 & 68 & 238 & 235 & 628 & 325 & 836 \\
        $0.63 - 0.72$ & 0.15 & 373 & 20 & 172 & 62 & 197 & 221 & 572 & 298 & 755 \\
        $0.72 - 0.8$ & 0.158 & 431 & 17 & 159 & 60 & 195 & 195 & 507 & 264 & 657 \\
        $0.8 - 0.9$ & 0.166 & 495 & 17 & 150 & 61 & 180 & 167 & 459 & 229 & 578 \\
        $0.9 - 1$ & 0.176 & 568 & 16 & 130 & 56 & 172 & 155 & 415 & 210 & 535 \\
        $1 - 1.1$ & 0.187 & 654 & 15 & 117 & 56 & 155 & 140 & 378 & 192 & 469 \\
        $1.1 - 1.22$ & 0.2 & 750 & 14 & 113 & 54 & 143 & 129 & 344 & 174 & 432 \\
        $1.22 - 1.41$ & 0.217 & 880 & 14 & 99 & 52 & 128 & 114 & 300 & 154 & 378 \\
        $1.41 - 2.5$ & 0.241 & 1068 & 12 & 80 & 48 & 107 & 83 & 229 & 110  & 285 \\
        $2.5 - 3.2$ & / & / & 10 & 69 & 46 & 98 & 62 & 161 & 80 & 195 \\
        $3.2 - 4.2$ & / & / & 10 & 62 & 46 & 86 & 51 & 131 & 66 & 158 \\
        $4.2 - 8$ & / & / & 8 & 49 & 43 & 74 & 38 & 102 & 49 & 122 \\
    \end{NiceTabular}
    }
    \tablefoot{Bin edges and corresponding values for the maximum wave number ($k_{\text{max}}^\text{NL}$, in $\rm Mpc$) and multipole ($\ell_{\text{max}}^\text{NL}$) adopted, based on restricting the analysis to linear/mildly nonlinear scales, and values of the best resolved angular scale $\ell_{\text{max}}^\text{LOC}$ and damping scale $\ell_{\text{damp}}$, coming from the finite sky resolution of GW events. This table refers to the ET $\Delta$ configuration.}
    \label{tab:valuesD}
\end{table*}

\begin{table*}[h]
    \caption{Bin edges for the ET 2L configuration.}
    \centering
        \resizebox{\textwidth}{!}{
    \begin{NiceTabular}{c||cc|cc|cc|cc|cc}
        & & & \multicolumn{2}{c}{ET 2L} & \multicolumn{2}{c}{ET 2L (loc. cut)} & \multicolumn{2}{c}{ET 2L + 1 CE} & \multicolumn{2}{c}{ET 2L + 2CE}\\
        \textbf{Bin Edges} & \textbf{$k_\text{max}^\text{NL}$} & \textbf{$\ell_\text{max}^\text{NL}$} & \textbf{$\ell_\text{damp}$} & \textbf{$\ell_\text{max}^\text{LOC}$} & \textbf{$\ell_\text{damp}$} & \textbf{$\ell_\text{max}^\text{LOC}$} & \textbf{$\ell_\text{damp}$} & \textbf{$\ell_\text{max}^\text{LOC}$} & \textbf{$\ell_\text{damp}$} & \textbf{$\ell_\text{max}^\text{LOC}$} \\
        \hline
        \hline
        $0 - 0.3$ & 0.117 & 130 & 165 & 718 & 179 & 732 & 877 & $>1000$ & 1200 & $>1000$ \\
        $0.3 - 0.38$ & 0.122 & 171 & 83 & 395 & 104 & 400 & 479 & $>1000$ & 634 & $>1000$ \\
        $0.38 - 0.47$ & 0.128 & 214 & 63 & 399 & 91 & 413 & 345 & $>1000$ & 471 & $>1000$\\
        $0.47 - 0.55$ & 0.135 & 263 & 52 & 287 & 73 & 308 & 327 & 822 & 445 & $>1000$ \\
        $0.55 - 0.63$ & 0.142 & 315 & 51 & 207 & 76 & 224 & 270 & 751 & 374 & 935 \\
        $0.63 - 0.72$ & 0.15 & 373 & 51 & 224 & 73 & 229 & 256 & 697 & 345 & 861 \\
        $0.72 - 0.8$ & 0.158 & 431 & 40 & 241 & 68 & 255 & 223 & 625 & 299 & 756 \\
        $0.8 - 0.9$ & 0.166 & 495 & 39 & 196 & 63 & 209 & 192 & 551 & 268 & 679 \\
        $0.9 - 1$ & 0.176 & 568 & 37 & 184 & 57 & 215 & 172 & 487 & 239 & 612 \\
        $1 - 1.1$ & 0.187 & 654 & 33 & 161 & 57 & 182 & 159 & 462 & 219 & 546 \\
        $1.1 - 1.22$ & 0.2 & 750 & 29 & 141 & 53 & 162 & 149 & 414 & 197 & 497 \\
        $1.22 - 1.41$ & 0.217 & 880 & 26 & 135 & 53 & 169 & 127 & 355 & 175 & 441 \\
        $1.41 - 2.5$ & 0.241 & 1068 & 20 & 105 & 49 & 135 & 94 & 268 & 125 & 324 \\
        $2.5 - 3.2$ & / & / & 15 & 87 & 49 & 116 & 70 & 190 & 92 & 226 \\
        $3.2 - 4.2$ & / & / & 13 & 78 & 48 & 110 & 58 & 154 & 75 & 180 \\
        $4.2 - 8$ & / & / & 9 & 58 & 46 & 89 & 43 & 117 & 56 & 138 \\
    \end{NiceTabular}
    }
    \tablefoot{As in Tab.~\ref{tab:valuesD}, for the ET 2L configuration.}
    \label{tab:values2L}
\end{table*}

\section{Tables}\label{app:tables}

We report here the values of the relative uncertainties obtained from various cases discussed in the main body of the paper.

Table~\ref{tab:snr} contains values of the S/N of the GW auto--correlation term as well as of the cross--correlation of GW catalogs with the Euclid photometric sample.
Table~\ref{tab:officialConstraints} contains constraints corresponding to the main configuration studied in the paper, namely the Euclid photometric survey in combination with 5 years of observations of different GW detector networks (see table), with 13 equally populated bins.
For comparison, we also report values for 1 and 10 years of observations for the configuration ET 2L + 2 CE in Tab.~\ref{tab:2L2CE_1_10_yrs}. 
Finally, Tables~\ref{tab:ETalone_separate}-\ref{tab:ET2CE_separate} contain the relative $1\sigma$ uncertainties obtained considering the auto and cross spectra of GWs and the Euclid photometric surveys separately.
\begin{table*}[t]
\caption{S/N.}
\centering
\begin{NiceTabular}{c||c|c}
Detector network & S/N GW-GW & S/N Cross-correlation\\
\hline
\hline
ET $\Delta$ & 0.06 & 1.28 \\
$\rm{ET\ 2L}$ & 0.2 & 3.7 \\
ET $\Delta$ + 1 CE & 1.2 & 15.4 \\
$\rm{ET\ 2L +\ 1\ CE}$ & 1.4 & 16.9 \\
ET $\Delta$ + 2 CE & 1.7 & 19.5 \\
$\rm{ET\ 2L +\ 2\ CE\ (T_{\rm obs}=1yr)}$ & 0.4 & 9.6 \\
$\rm{ET\ 2L +\ 2\ CE\ (T_{\rm obs}=5yr)}$ & 1.8 & 21.2\\
$\rm{ET\ 2L +\ 2\ CE\ (T_{\rm obs}=10yr)}$ & 3.5 & 29.6\\
\end{NiceTabular}%
\tablefoot{S/N of the auto--correlation of GW sources and of their cross--correlation with the Euclid photometric sample, for different GW detector configurations.}
\label{tab:snr}
\end{table*}
\begin{table*}[t]
\caption{Relative uncertainties.}
\centering
\resizebox{\textwidth}{!}{%
\begin{NiceTabular}{c||c|c|c|c|c|c}
    Parameter & ET $\Delta$ & ET 2L & ET $\Delta$ + 1 CE & ET 2L + 1 CE & ET $\Delta$ + 2 CE & ET 2L + 2 CE \\
    \hline
    \hline
    $H_0$ & 8.7 & 5.8 & 1.5 & 1.3 & 1.2 & 1.1\\
    $\Omega_\mathrm{m}$ & 6.9 & 6.5 & 6.2 & 6.2 & 6.1 & 6.1\\
    $\Omega_\mathrm{b} h^2$ & 29.6 & 20.0 & 7.6 & 7.3 & 7.0 & 6.9\\
    $A_s$ & 12.4 & 10.2 & 8.3 & 8.3 & 8.2 & 8.2\\
    $n_s$ & 7.1 & 5.7 & 4.3 & 4.3 & 4.2 & 4.2\\
    $b_\mathrm{g}$ bin 1-5 & 6.1 & 5.7 & 5.3 & 5.3 & 5.3 & 5.3\\
    $b_\mathrm{g}$ bin 6-10 & 5.9 & 5.5 & 5.3 & 5.3 & 5.3 & 5.3\\
    $b_\mathrm{g}$ bin 11-13 & 5.9 & 5.5 & 5.3 & 5.3 & 5.3 & 5.3\\
    $b_{\rm GW}$ bin 1-5 & $>100$ & $>100$ & 60 & 60 & 40 & 40\\
    $b_{\rm GW}$ bin 6-10 & $\gg 100$ & $\gg 100$ & 55 & 50 & 35 & 35\\
    $b_{\rm GW}$ bin 11-16 & $\gg 100$ & $\gg 100$ & $>100$ & $>100$ & $>100$ & 90\\
\end{NiceTabular}
}
\tablefoot{Relative $1\sigma$ uncertainty on all the parameters of the analysis, for the Euclid photometric survey in combination with 5 years of observations of different GW detector networks with 13 equally populated bins.}
\label{tab:officialConstraints}
\end{table*}
\begin{table*}[t]
\caption{Relative uncertainties with different observing times.}
\centering
\resizebox{\textwidth}{!}{%
\begin{NiceTabular}{c||c|c|c|c||c|c|c|c}
    ET 2L + 2 CE & \multicolumn{4}{c}{$\mathrm{T_{obs}=1}$ years} & \multicolumn{4}{c}{$\mathrm{T_{obs}=10}$ years} \\
    \hline
    \hline
    Parameter & Total & Galaxy & GW & Cross & Total & Galaxy & GW & Cross\\
    \hline
    $H_0$ & 1.7 & 15.2 & $>100$ & 16.4 & 0.9 & 15.2 & 27 & 5.1 \\
    $\Omega_m$ & 6.2 & 8.2 & $\gg 100$ & $>100$ & 6.0 & 8.2 & $>100$ & 44.6\\
    $\Omega_b h^2$ & 8.3 & 23.3 & $\gg 100$ & $\gg 100$ & 6.7 & 23.3 & $\gg 100$ & $>100$\\
    $A_s$ & 8.3 & 18.1 & $\gg 100$ & $\gg 100$ & 8.2 & 18.1 & $\gg 100$ & $>100$\\
    $n_s$ & 4.3 & 10.8 & $\gg 100$ & $>100$ & 4.2 & 10.8 & $\gg 100$ & 43\\
    $b_g$ bin 1-5 & 5.4 & 6.8 & / & $\gg 100$ & 5.3 & 6.8 & / & $>100$\\
    $b_g$ bin 6-10 & 5.4 & 6.9 & / & $\gg 100$ & 5.3 & 6.9 & / & 90\\
    $b_g$ bin 11-13 & 5.4 & 6.9 & / & $\gg 100$ & 5.3 & 6.9 & / & $>100$\\
    $b_{GW}$ bin 1-5 & $>100$ & / & $\gg 100$ & $\gg 100$ & 30 & / & $\gg 100$ & $>100$\\
    $b_{GW}$ bin 6-10 & 80 & / & $\gg 100$ & $\gg 100$ & 25 & / & $\gg 100$ & 80\\
    $b_{GW}$ bin 11-16 & $>100$ & / & $\gg 100$ & $\gg 100$ & 40 & / & $\gg 100$ & $>100$ \\
\end{NiceTabular}%
}
\tablefoot{Relative $1\sigma$ uncertainty on all the parameters of the analysis, for the Euclid photometric survey in combination with 1 and 10 years of observations of ET 2L $15\ \rm km\ + 2\ CE$ with 13 equally populated bins.}
\label{tab:2L2CE_1_10_yrs}
\end{table*}
\begin{table*}[t]
\caption{Relative uncertainties for ET alone.}
\centering
\resizebox{\textwidth}{!}{%
\begin{NiceTabular}{c||c|c|c|c||c|c|c|c}
    & \multicolumn{4}{c}{ET $\Delta$} & \multicolumn{4}{c}{ET 2L} \\
    \hline
    \hline
    Parameter & Total & Galaxy & GW & Cross & Total & Galaxy & GW & Cross \\
    \hline
    $H_0$ & 8.7 & 15.2 & $\gg 100$ & 73 & 5.8 & 15.2 & $\gg 100$ & 41 \\
    $\Omega_m$ & 6.9 & 8.2 & $\gg 100$ & $\gg 100$  & 6.5 & 8.2 & $\gg 100$ & $\gg 100$ \\
    $\Omega_b h^2$ & 29.6 & 23.3 & $\gg 100$ & $\gg 100$  & 20.0 & 23.3 & $\gg 100$ & $\gg 100$ \\
    $A_s$ & 12.4 & 18.1 & $\gg 100$ & $\gg 100$  & 10.2 & 18.1 & $\gg 100$ & $\gg 100$ \\
    $n_s$ & 7.1 & 10.8 & $\gg 100$ & $\gg 100$  & 5.7 & 10.8 & $\gg 100$ & $\gg 100$\\
    $b_g$ bin 1-5 & 6.1 & 6.8 & / & $\gg 100$ & 5.7 & 6.8 & / & $\gg 100$ \\
    $b_g$ bin 6-10 & 5.9 & 6.9 & / & $\gg 100$  & 5.5 & 6.9 & / & $\gg 100$ \\
    $b_g$ bin 11-13 & 5.9 & 6.9 & / & $\gg 100$  & 5.5 & 6.9 & / & $\gg 100$ \\
    $b_{GW}$ bin 1-5 & $>100$ & / & $\gg 100$ & $\gg 100$ & $>100$ & / & $\gg 100$ & $\gg 100$ \\
    $b_{GW}$ bin 6-10 & $\gg 100$ & / & $\gg 100$ & $\gg 100$  & $\gg 100$ & / & $\gg 100$ & $\gg 100$ \\
    $b_{GW}$ bin 11-16 & $\gg 100$ & / & $\gg 100$ & $\gg 100$  & $\gg 100$ & / & $\gg 100$ & $\gg 100$ \\
\end{NiceTabular}%
}
\tablefoot{Separate and total contributions of the galaxy-galaxy, GW-galaxy and GW-GW auto and cross--correlation terms to the constraints on cosmological and bias parameters, for the ET detectors alone in combination with the Euclid photometric survey, with 13 equally populated bins.}
\label{tab:ETalone_separate}
\end{table*}

\begin{table*}[t]
\caption{Relative uncertainties for ET in combination with one CE.}
\centering
\resizebox{\textwidth}{!}{%
\begin{NiceTabular}{c||c|c|c|c||c|c|c|c}
    & \multicolumn{4}{c}{ET $\Delta$ + 1 CE} & \multicolumn{4}{c}{ET 2L + 1 CE} \\
    \hline
    \hline
    Parameter & Total & Galaxy & GW & Cross  & Total & Galaxy & GW & Cross\\
    \hline
    $H_0$ & 1.5 & 15.2 & 95 & 12.1 & 1.3 & 15.2 & 85 & 11 \\
    $\Omega_m$ & 6.2 & 8.2 & $\gg 100$ & 91 & 6.2 & 8.2 & $\gg 100$ & 83.2\\
    $\Omega_b h^2$ & 7.6 & 23.3 & $\gg 100$ & $>100$ & 7.3 & 23.3 & $\gg 100$ & $>100$ \\
    $A_s$ & 8.3 & 18.1 & $\gg 100$ & $>100$  & 8.3 & 18.1 & $\gg 100$ & $>100$ \\
    $n_s$ & 4.3 & 10.8 & $\gg 100$ & 95 & 4.3 & 10.8 & $\gg 100$ & 87 \\
    $b_g$ bin 1-5 & 5.3 & 6.8 & / & $\gg 100$ & 5.3 & 6.8 & / & $>100$ \\
    $b_g$ bin 6-10 & 5.3 & 6.9 & / & $\gg 100$  & 5.3 & 6.9 & / & $>100$ \\
    $b_g$ bin 11-13 & 5.3 & 6.9 & / & $\gg 100$  & 5.3 & 6.9 & / & $>100$ \\
    $b_{GW}$ bin 1-5 & 60 & / & $\gg 100$ & $\gg 100$ & 60 & / & $\gg 100$ & $>100$ \\
    $b_{GW}$ bin 6-10 & 55 & / & $\gg 100$ & $\gg 100$  & 50 & / & $\gg 100$ & $>100$ \\
    $b_{GW}$ bin 11-16 & $>100$ & / & $\gg 100$ & $\gg 100$ & $>100$ & / & $\gg 100$ & $>100$ \\
\end{NiceTabular}%
}
\tablefoot{As in Tab.~\ref{tab:ETalone_separate}, for ET in combination with one CE.}
\end{table*}

\begin{table*}[t]
\caption{Relative uncertainties for ET in combination with two CE.}
\centering
\resizebox{\textwidth}{!}{%
\begin{NiceTabular}{c||c|c|c|c||c|c|c|c}
    & \multicolumn{4}{c}{ET $\Delta$ + 2 CE} & \multicolumn{4}{c}{ET 2L + 2 CE} \\
    \hline
    \hline
    Parameter & Total & Galaxy & GW & Cross & Total & Galaxy & GW & Cross \\
    \hline
    $H_0$ & 1.2 & 15.2 & 64 & 8.2 & 1.1 & 15.2 & 54 & 7.2 \\
    $\Omega_m$ & 6.1 & 8.2 & $\gg 100$ & 64.5 & 6.1 & 8.2 & $\gg 100$ & 54\\
    $\Omega_b h^2$ & 7.0 & 23.3 & $\gg 100$ & $>100$ & 6.9 & 23.3 & $\gg 100$ & $>100$ \\
    $A_s$ & 8.2 & 18.1 & $\gg 100$ & $>100$ & 8.2 & 18.1 & $\gg 100$ & $>100$ \\
    $n_s$ & 4.2 & 10.8 & $\gg 100$ & 72 & 4.2 & 10.8 & $\gg 100$ & 64 \\
    $b_g$ bin 1-5 & 5.3 & 6.8 & / & $>100$ & 5.3 & 6.8 & / & $>100$ \\
    $b_g$ bin 6-10 & 5.3 & 6.9 & / & $>100$ & 5.3 & 6.9 & / & $>100$ \\
    $b_g$ bin 11-13 & 5.3 & 6.9 & / & $>100$ & 5.3 & 6.9 & / & $>100$ \\
    $b_{GW}$ bin 1-5 & 40 & / & $\gg 100$ & $>100$ & 40 & / & $\gg 100$ & $>100$ \\
    $b_{GW}$ bin 6-10 & 35 & / & $\gg 100$ & $>100$ & 35 & / & $\gg 100$ & $>100$ \\
    $b_{GW}$ bin 11-16 & $>100$ & / & $\gg 100$ & $>100$ & 90 & / & $\gg 100$ & $>100$ \\
\end{NiceTabular}%
}
\tablefoot{As in Tab.~\ref{tab:ETalone_separate}, for ET in combination with two CE.}
\label{tab:ET2CE_separate}
\end{table*}